\documentclass[pdflatex,sn-mathphys-num]{sn-jnl}% Math and Physical Sciences Numbered Reference Style 
\usepackage{graphicx}%
\usepackage{multirow}%
\usepackage{amsmath,amssymb,amsfonts}%
\usepackage{amsthm}%
\usepackage{mathrsfs}%
\usepackage[title]{appendix}%
\usepackage{xcolor}%
\usepackage{textcomp}%
\usepackage{manyfoot}%
\usepackage{booktabs}%
\usepackage{algorithm}%
\usepackage{algorithmicx}%
\usepackage{algpseudocode}%
\usepackage{listings}%
\usepackage{subcaption}%
\usepackage{placeins}%
\usepackage[utf8]{inputenc}%
\theoremstyle{thmstyleone}%
\theoremstyle{thmstyletwo}%

\theoremstyle{thmstylethree}%

\DeclareUnicodeCharacter{0300}{\`{}} % or the appropriate command for the accent

\begin{document}

\title[Article Title]{GraphAge: Decoding Epigenetic Aging through Co-Methylation Network Analysis Using GNN and GNN explainer}

\title[GraphAge]{GraphAge: Unleashing the power of Graph Neural Network to Decode Epigenetic Aging}


%%%for double blind review the following is commented.
\author[1]{\fnm{Saleh Sakib} \sur{Ahmed}}\email{birdhunterx91@gmail.com}

\author[1]{\fnm{Nahian} \sur{Shabab}}\email{nsnbd134@gmail.com}

\author[2]{\fnm{Md. Abul Hassan } \sur{Samee}}\email{samee@bcm.edu}

\author*[1]{\fnm{M. Sohel } \sur{Rahman}}\email{sohel.kcl@gmail.com}

\affil*[1]{\orgdiv{Computer Science and Engineering}, \orgname{Bangladesh University of Engineering and Technology}, \orgaddress{\street{West Palashi}, \city{Dhaka}, \postcode{1205}, \state{Dhaka}, \country{Bangladesh}}}

\affil[2]{\orgdiv{Integrative Physiology}, \orgname{Baylor College of Medicine}, \orgaddress{\city{Houston}, \state{TX}, \country{USA}}}

%%%for double blind review the above is commented.

\abstract{

%\subsection{Background}
DNA methylation is a crucial epigenetic marker used in various clocks to predict epigenetic age. However, many existing clocks fail to account for crucial information about CpG sites and their interrelationships, such as co-methylation patterns.
%\subsection{Methods}
We present a novel approach to represent methylation data as a graph, using methylation values and relevant information about CpG sites as nodes, and relationships like co-methylation, same gene, and same chromosome as edges. We then use a Graph Neural Network (GNN) to predict age. Thus our model, GraphAge  leverages both the structural and positional information for prediction as well as better interpretation.
%\subsection{Results}
Although, we had to train in a constrained compute setting, GraphAge still showed competitive performance with a Mean Absolute Error (MAE) of ~3.207 and a Mean Squared Error (MSE) of ~25.277, slightly outperforming the current state of the art. Perhaps more importantly, 
%compared to AltumAge's MAE of ~3.296 and MSE of ~28.310. 
we utilized GNN explainer for interpretation purposes and were able to unearth interesting insights (e.g., key CpG sites, pathways and their relationships through Methylation Regulated Networks in the context of aging), which were not  possible to `decode' without leveraging the unique capability of GraphAge to `encode' various structural relationships. 
%importance of nodes (CpG sites), edges (their relationships) and node attributes (various biological information).
%\subsection{Conclusion}
GraphAge has the potential to consume and utilize all relevant information (if available) about an individual that relates to the complex process of aging. So, in that sense it is one of its kind and can be seen as the first benchmark for a multimodal model which can incorporate all these information in order to close the gap in our understanding of the true nature of aging. 
%We identified key CpG sites and relationships, forming subnetworks that we termed Methylation Regulated Networks. Additionally, we identified hypermethylating and hypomethylating pathways. We were also able to establish a relationships among these two types of pathways.Also with the help of attribute importance we were able to understand the contribution of various biological information to aging.
% CpG site methylation, the addition of a methyl group to specific DNA sites, regulates gene expression. When these sites are hypomethylating (less methylation), gene expression increases. Conversely, hypermethylation (more methylation) decreases gene expression. CpG sites can also influence each other's methylation through co-methylation, affecting gene regulation and aging. Understanding these interactions is crucial for studying epigenetic aging. So we propose a novel, scalable approach to the methylation problem by representing CpG site information as nodes and their relationships as edges in a graph. By leveraging the power of GNNs, we achieved competitive results. Additionally,using GNN explainer, we identified the importance of different CpG sites and their relationships in those graphs. After filtering, we identified critical subnetworks of important connected CpG sites, we termed as Methylation Regulated Networks (MRNs). Additionally, incorporating other relevant biological information of CpG sites into the nodes allowed us to determine their importance using the GNN explainer.
}




\maketitle

\section{Introduction}\label{intro}

The pursuit of a cure for aging has fascinated humanity for centuries, symbolized by the mythical quests for the Fountain of Youth or Elixir of Life. In contemporary times, these quests have transformed into scientific endeavors, with researchers exploring the intricate biology of aging and the underlying mechanisms that drive it. Modern studies focus on understanding how molecular processes and/or elements, such as DNA methylation, Gene expression, Glycans, proteomics and other biomarkers, influence aging and longevity \cite{jylhava2017biological,lopez2013hallmarks}. All these are information stored in cells and aging is a result from the loss of regulatory information in cells, leading to the accumulation of epigenetic noise and cellular dysfunction \cite{lu2023information}. But by deciphering these processes, scientists aim to uncover potential interventions that could delay or even reverse aspects of biological aging \cite{tarkhov2022aging,fahy2019reversal}.

This ongoing research into the epigenetic regulation of aging is not just a continuation of an age-old quest, but a sophisticated, data-driven effort to map the biological pathways that govern the aging process. Key to this exploration is the role of CpG sites and their influence on gene regulation through methylation patterns \cite{bell2011dna,jaenisch2003epigenetic,ladd2015epigenetic}, which hold significant implications on our understanding of chronological age and healthspan \cite{marsit2015influence,gomes2013epigenetic,mckay2011diet,godfrey2015developmental,obata2015epigenetic}. 

Various approaches to modeling the aging process have been explored in the literature, including transcriptomic models \cite{holzscheck2021modeling,peters2015transcriptional}, which have demonstrated great generalizability. Glycans are also considered good biomarkers for age \cite{krivstic2014glycans}, and glycan clocks have shown some merit \cite{mijakovac2022heritability}. Proteomics clocks have also been utilized. These models represent interconnected information, but unfortunately, there is no unified dataset where all information of a single individual is present. Among all these approaches, epigenetic clocks have proven to be the most influential \cite{jylhava2017biological}.

\subsection{Prior Works}
%%%%%%%%%%%%%%%%%%%%%%%%%  PLEASE CHECK %%%%%%%%%%%%%%%%%%%%%%%%%%%%%%%%%%%%%%%%%%%%%%%%5
%DNA methylation, the addition of a methyl group to the C5 position of cytosine to form 5-methylcytosine and they predominantly occur at CpG sites \cite{moore2013dna, jabbari2004cytosine}. This biochemical process is crucial in regulating gene transcription. hypomethylation, or the decrease in methylation levels, leads to the upregulation (increase) of gene transcription, while hypermethylation, or the increase in methylation levels, results in the downregulation (decrease) of gene transcription \cite{attwood2002dna}.

Early research has unveiled a profound link between DNA methylation and aging, highlighting its significant role in the aging process \cite{boks2009relationship, richardson2003impact}. This groundbreaking insight spurred researchers to develop the first epigenetic clock, a remarkable tool created using saliva samples in 2011 \cite{bocklandt2011epigenetic}. This innovation marked a pivotal moment in epigenetics. Subsequently, Hannum et al. \cite{hannum2013genome} published their 71 site methylation clock trained on blood followed by Horvath's seminal work on a pan-tissue epigenetic clock on 353 CpG sites, demonstrating that the chronological age of an organism can be estimated based on epigenetic modifications, particularly DNA methylation patterns. Unlike chronological age, which simply counts the number of years a person has lived, epigenetic age aims to reflect the biological condition of the body and its systems, potentially providing a more accurate indicator of aging and health status. Many subsequent studies have aimed to refine and enhance these predictions (e.g., \cite{galkin2021deepmage,levy2020methylnet,ying2024causality}). Horvath's model, using linear regression on DNA methylation data, provided a robust framework for estimating chronological ages with impressive accuracy \cite{horvath2013dna}.

In 2021, Galkin et. al. \cite{galkin2021deepmage} introduced DeepMAge, a deep neural network model that improved prediction performance over Horvath’s original model, particularly in blood samples. They employed gradient-based feature selection followed by a sequential neural network with four hidden layers, each containing 512 neurons. They used pathway analysis on the 1000 selected CpG sites by feature selection. They also emphasized the need for multimodal dataset, a lacking that still prevails as of today. %which their model lacked. 
As will be clear later, despite the current lack of such comprehensive datasets, our proposed model shows promise as a multimodal model.

Similarly, Levy et. al. \cite{levy2020methylnet} achieved encouraging results in age prediction using a multi-layer perceptrons. Another significant development, AltumAge \cite{lapierre2022pan}, employed multi-layer perceptron layers and experimented their model performance across multiple tissues. They further utilized SHAP (Shapley Additive Explanations) \cite{lundberg2017unified} to interpret the contributions of different CpG sites towards age prediction. AltumAge also used interaction scores obtained from SHAP to demonstrate relationships between hyper-and hypomethylating CpG sites. Note that, hypomethylation and hypermethylation of DNA refer to relatively less or more methylation than in some standard DNA. %(i.e., increased addition of methyl groups to the DNA molecule) and hypomethylating (i.e., decreased addition of methyl groups to the DNA molecule) CpG sites. 
%Note that hypomethylation refers to a decrease in DNA methylation levels, while hypermethylation refers to an increase. These changes can occur in comparison to the levels typically observed at a younger age.
Additionally, they identified pathways involving their top CpG sites. Thus the authors attempted to interpret and explain the results instead of just using AltmAge as a black-box. As will be shown later we also take a big stride along this dimension and hence we use AltumAge as the current state of the art for benchmarking. 

Ying et. al. \cite{ying2024causality} proposed using epigenome-wide Mendelian Randomization \cite{relton2015mendelian} (EWMR) on 420,509 CpG sites to identify the sites that are causal to twelve aging-related traits, and then running the selected sites through an elastic net regression \cite{zou2005regularization} model. A notable concern raised by Ying et al. was that correlation, rather than causality, dominated the results due to the exclusive use of methylation data. Therefore, they focused on using other biological data about the CpG sites for feature selection to achieve a more biologically significant outcome. This brings up the question of whether we could have a generalized formulation that would incorporate any additional feature (e.g., the ones mentioned by Ying et. al.) in the model under consideration, assuming of course, that the data of all relevant sites are available. As will be clear later, our model, GraphAge, naturally lends itself to such a formulation. %and based on these additional information for feature selection be used to make prediction. As you will see through out the paper our model is a suitable candidate.

%But what if there was a generalized formulation that could easily incorporate any additional feature into its feature space the only thing required will be availability of data of all relevant sites then I think all concerns would be satisfied 

% So, to summarize, epigenetic clocks in the literature are either pan-tissue\cite{horvath2013dna,lapierre2022pan} or tissue specific \cite{galkin2021deepmage,ying2024causality,simpson2021epigenetic} and they all have a feature selection mechanism, some are as simple as taking common CpG across multiple platforms \cite{lapierre2022pan}, others using elastic net \cite{galkin2021deepmage} and/or more complicated methods \cite{ying2024causality}; but they all aim to answer two questions: (a) What is the epigenetic age, and (b) what are the relevant pathways? 

Before concluding this section, a brief discussion on a related interesting concept, called co-methylation is in order. Affinito et al. \cite{affinito2020nucleotide} showed that co-methylation depends on spatial and structural information and is related to the nucleotide distance among CpG sites. The co-methylation of closer CpG sites have more influence than that of those that are further apart. Also, research findings indicate that after DNA replication, De Novo genes, depending on specific sites,  may or may not exhibit the same methylation rates as their parents, suggesting that methylation is influenced by structural and positional information \cite{genereux2005population}. This has motivated the researchers to model co-methylation networks along with constructing models to predict aging. For example, Wu et al. \cite{wu2019m6acomet} created a co-methylation network based on MeRIP-Seq data by taking correlations among all sites and visualizing it using Cytoscape \cite{shannon2003cytoscape}.

% <First you need to add a line to connect methylation with epigenetic aging here. Otherwise this seems unrelated>. Consequently, along with constructing models for predicting aging, we notice several attempts in the literature to model the co-methylation network. For example, Wu et al. \cite{wu2019m6acomet} created a co-methylation network based on MiSeq data by taking correlations among all sites and visualizing it using Cytoscape. Affinito et al. \cite{affinito2020nucleotide} showed that co-methylation depends on spatial and structural information and that co-methylation had a relationship with distance of CpG sites \cite{affinito2020nucleotide}. Also it was shown that after replication De Novo methylation might not be methylating depending on position \cite{genereux2005population}.

\subsection{Our Contributions}     

All previous clock models had fairly good accuracy but they did not adequately take into account crucial structural information and the dynamic interactions between CpG sites through co-methylation. This limits our understanding of interaction among these sites thereby preventing us from achieving a comprehensive interpretation. On the other hand, works that tried to model the co-methylation network did not take into account how CpG sites influence the aging process and which co-methylation edges/CpG sites are important to give a meaningful interpretation. 

We propose to solve both the problems by utilizing all possible given information we currently have available regarding the CpG sites to formulate a graph where nodes encapsulate the information about CpG sites and edges represent all relationships among them thereby capturing the structure of co-methylation network. 
%. As a result, it considers the structure of co-methylation network then finally
Then, we train it using the power of Graph Neural Networks, albeit in a resource constraint setting. Thus, we present the first age prediction model, GraphAge, that attempts to leverage all available structural information \cite{zhou2020graph}. And despite our resource constraints, GraphAge was able to ``come of age" by outperforming AltumAge in both dimensions of efficacy and interpretatbility. Notably, instead of a pan-tissue training approach, we opted for a tissue specific approach because gene expression follows a tissue specific pattern \cite{sonawane2017understanding,ong2011enhancer}. For interpretation purposes, we leverage the Graph Neural Network Explainer \cite{ying2019gnnexplainer}. 
%to give us sites that are important as well as edges that are important which we use to make our Methylation Regulated Network through filtering. Then we made a very visual representation where the researchers can have a very good understanding of how the sites operate.

%And for choosing between tissue specific training or pan tissue we chose the former. Because while there is nothing preventing us from using a pan tissue  approach we opted for a tissue specific approach because gene expression follows a tissue specific pattern \cite{sonawane2017understanding,ong2011enhancer} and that is why we trained our model in a tissue specific way.
\begin{figure}[h!]
    \centering
    \begin{subfigure}[b]{\textwidth}
        \centering
        \includegraphics[width=\textwidth]{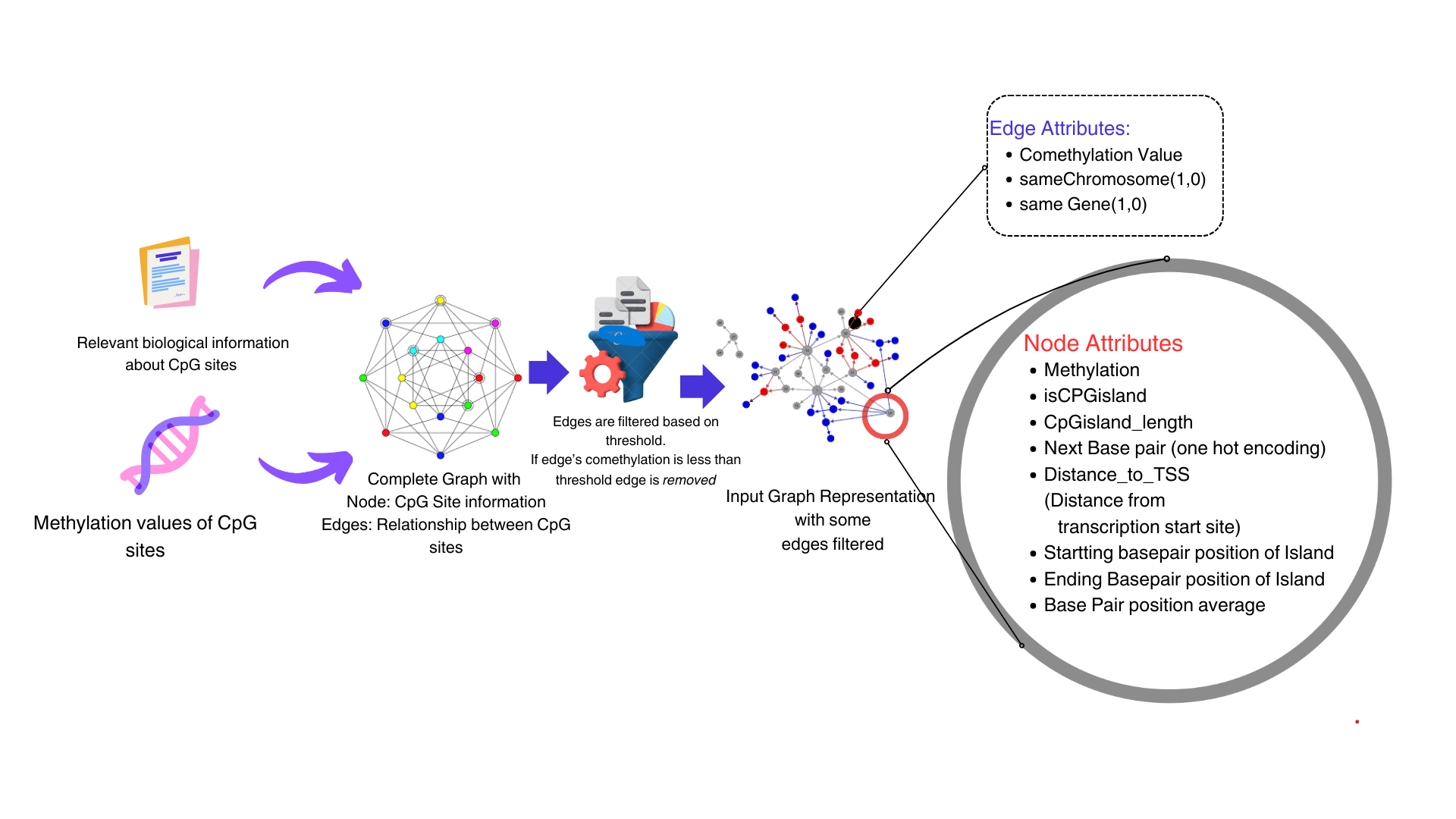}
        \caption{}
        \label{fig:2a}
    \end{subfigure}
    \begin{subfigure}[b]{\textwidth}
        \centering
        \includegraphics[width=\textwidth]{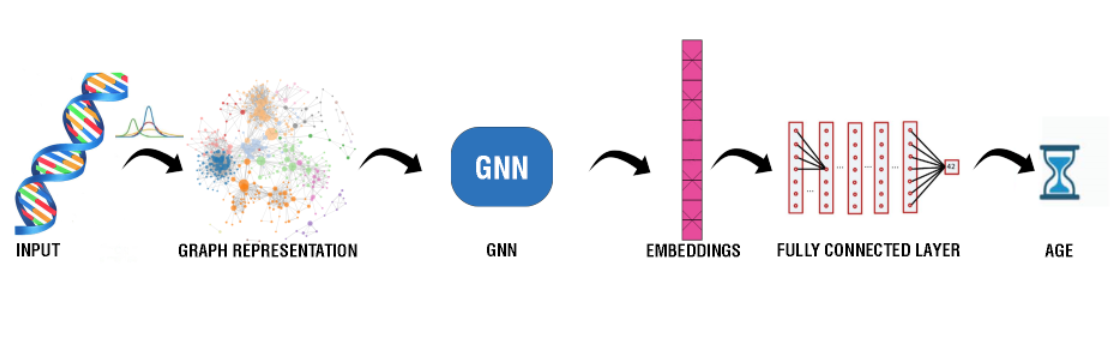}
        \caption{}
        \label{fig:2b}
    \end{subfigure}
    \caption{\textbf{a.} We possess tabular methylation data from various platforms with CpG site information from supplementary data. This information was transformed into a graph, where node attributes represent CpG site information and edge attributes represent their relationships. We calculated co-methylation values among all sites and filtered out edges below a certain threshold. We applied a universal threshold and chromosome-specific thresholds. The chromosome-specific thresholds permit the inclusion of lower-value edges within the same chromosome, with an even lower threshold for edges within a certain distance on the same chromosome. Finally, we obtained our graph representation as inputs. \textbf{b.}This diagram illustrates our process flow. We convert tabular data into graph representations and feed them into a PNA GNN (a graph convolution layer) to obtain an embedding. This embedding is then passed through a fully connected layer to generate the final age prediction.  }
    \label{fig:2}
\end{figure}

\section{Results and Discussions}\label{sec2}
\subsection{A brief Methodical Overview}
For the sake of better comprehension, here we give a brief overview of our methods (Fig. \ref{fig:2}); the details can be found in Section \ref{sec:method}. We constructed a graph where nodes are CpG sites and edges represent their relationships. We selected CpG sites common across all methylation data platforms, using structural information from NCBI GEO \cite{ncbi_geo}. We transformed tabular methylation data into a graph, with node attributes including methylation beta values, CpG island information, base pair positions, and normalized distances from transcription start sites. Edge features included co-methylation values, same chromosome, and gene indicators. We applied universal, chromosome-specific, and distance-based thresholds for edge filtering, allowing more flexibility for closer CpG sites with lower co-methylation values \cite{affinito2020nucleotide}. Our model used a Principal Neighborhood Aggregation (PNA) \cite{corso2020principal} convolutional layer in a Graph Neural Network (GNN) followed by a fully connected layer for age prediction.

For interpreting the model, we used GNN explainer \cite{ying2019gnnexplainer}. We averaged node and edge importance across age groups for a comprehensive understanding and visualized these using Graphviz \cite{graphviz}. Importance analysis involved averaging node attribute importance across samples and plotting temporal patterns. We performed Enrichr analysis \cite{chen2013enrichr,kuleshov2016enrichr,xie2021gene} on hypo- and hypermethylating genes, and identified nodes with significant upward or downward trends using linear regression.
% \FloatBarrier
 \begin{figure}[htbp!]
    \centering
    % INDEPENDENT SET
    \begin{subfigure}[b]{0.26\textwidth}
        \begin{subfigure}[b]{\textwidth}
        \centering
         \includegraphics[width=0.8\textwidth]{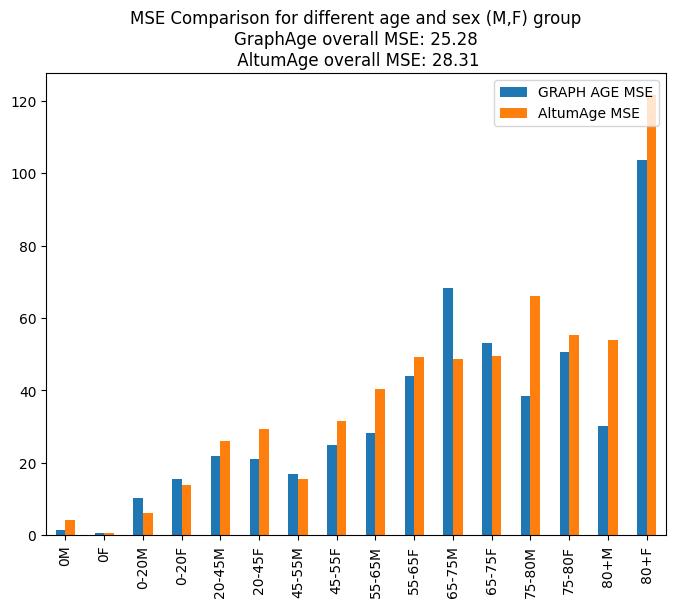}
        \caption{}
        \label{fig:mse}
        
        \end{subfigure}

        \begin{subfigure}[b]{\textwidth}
            \centering
             \includegraphics[width=0.8\textwidth]{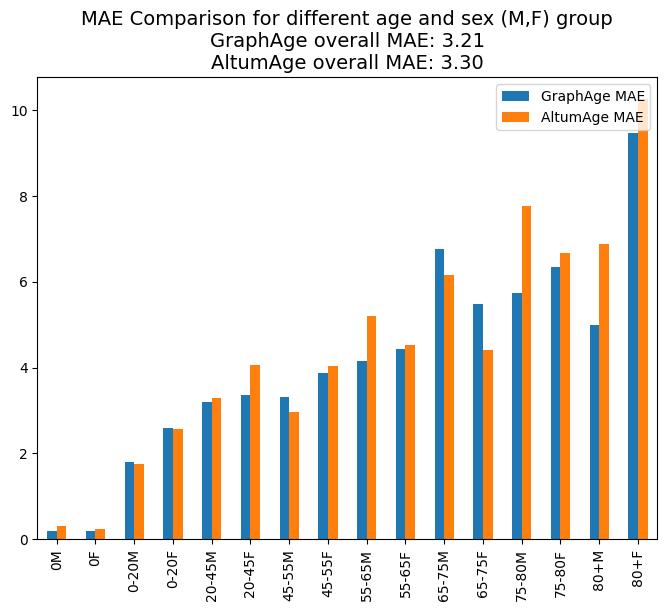}
            \caption{}
            \label{fig:mae}
        \end{subfigure}
        % \caption{}
        \label{independet set performance}
    \end{subfigure}
    % Case Study
    \begin{subfigure}[b]{0.25\textwidth}
        \begin{subfigure}[b]{\textwidth}
            \centering
            \includegraphics[width=\textwidth]{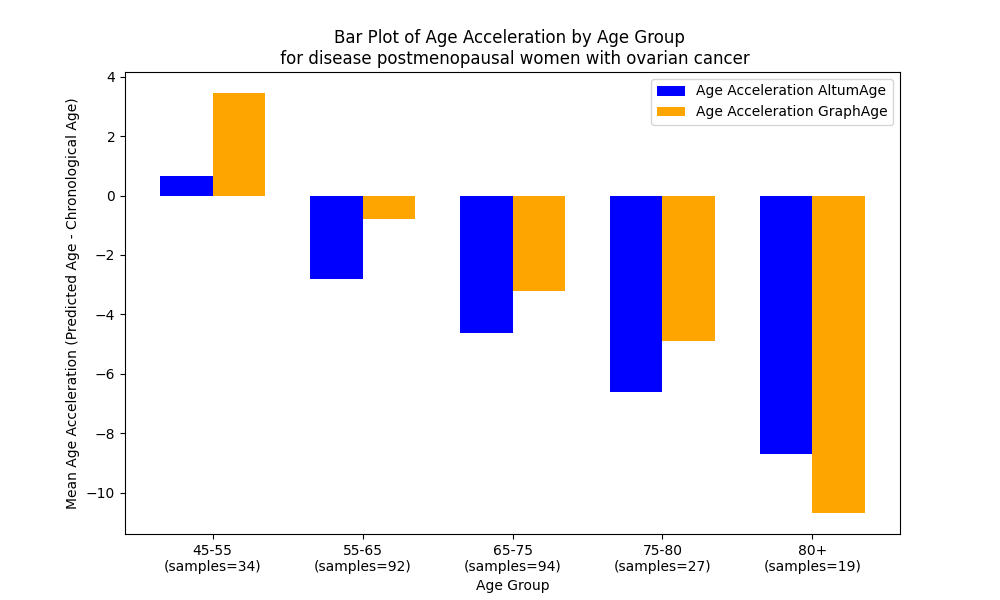}
            \label{fig:overian cancer}
        \end{subfigure}
        \begin{subfigure}[b]{\textwidth}
            \centering
            \includegraphics[width=\textwidth]{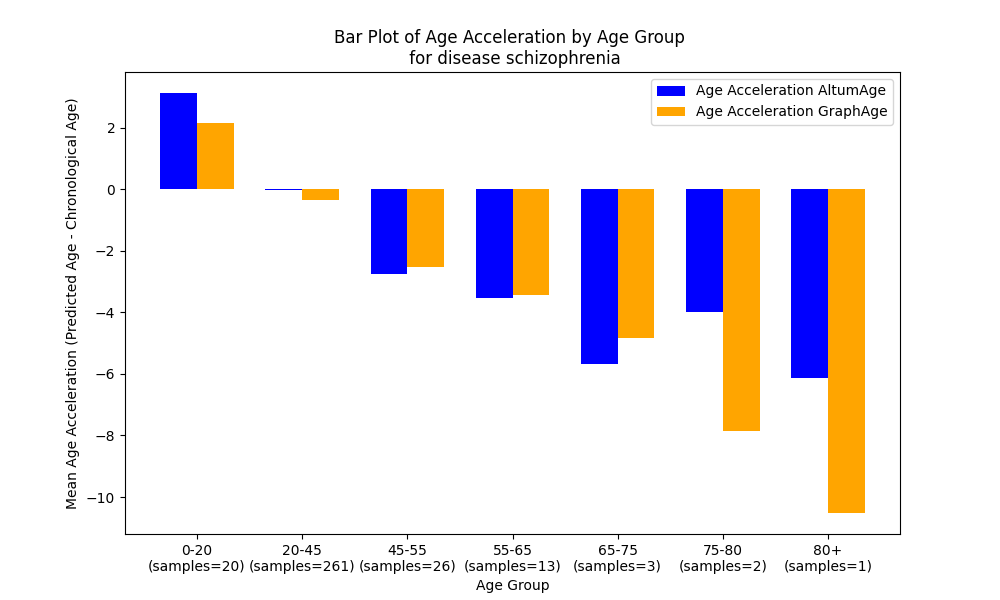}
            \label{fig:schrizophrania}
        \end{subfigure}
        \begin{subfigure}[b]{\textwidth}
        \centering
        \includegraphics[width=\textwidth]{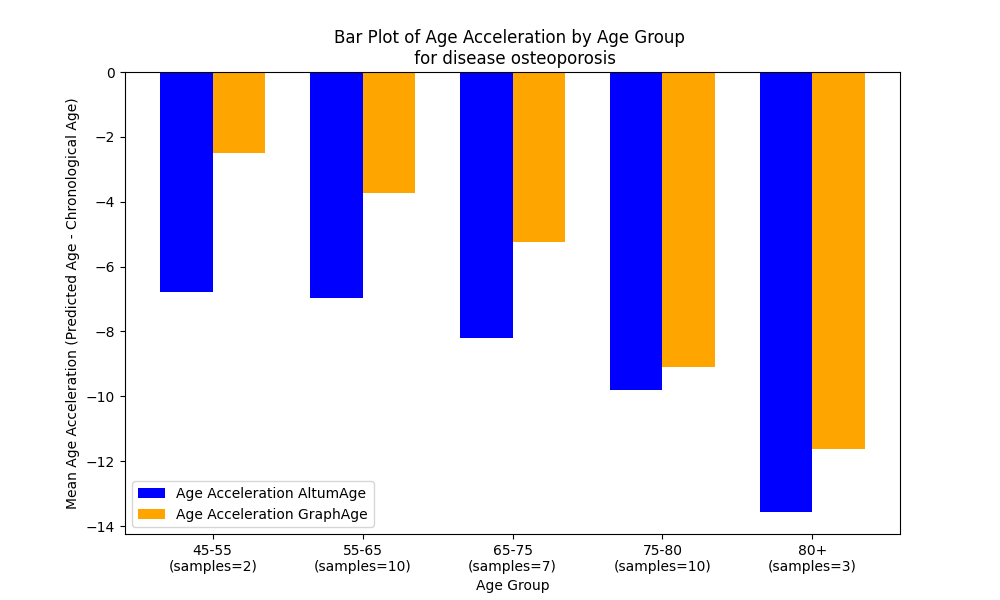}
        
        \label{fig:osteoporosis}
        \end{subfigure}
        \caption{}
        \label{case study performance}
    \end{subfigure}
    \begin{subfigure}[b]{0.26\textwidth}
        \begin{subfigure}[b]{\textwidth}
        \centering
        \includegraphics[width=\textwidth]{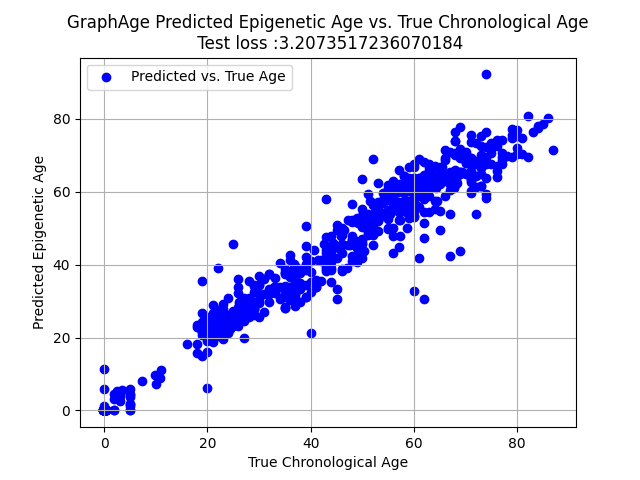}
        
        \label{fig:1b1}
        \end{subfigure}
        \begin{subfigure}[b]{\textwidth}
            \centering
            \includegraphics[width=\textwidth]{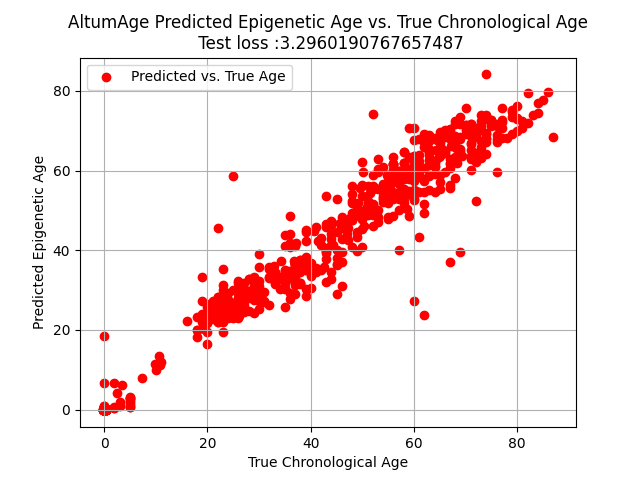}
            \caption{}
        \label{bio_vs_pred}
        \end{subfigure}
    \end{subfigure}
    \begin{subfigure}[b]{0.65\textwidth}
            \centering
            \includegraphics[width=0.5\textwidth]{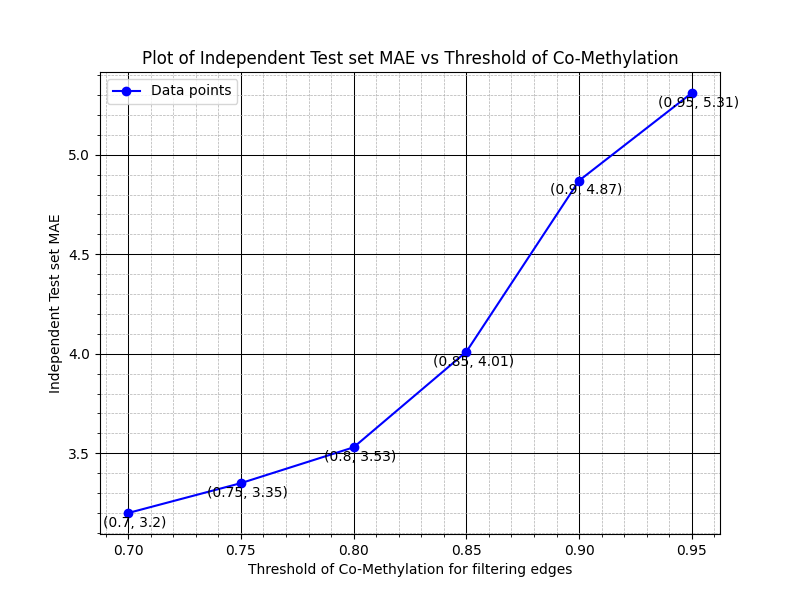}
            
            \caption{}
    \label{threshold vs error}

    \end{subfigure}
    
    \caption{\textbf{a. \& b.} We trained our model using 5-fold cross-validation and then tested it on the test set. %The performance of the model, with a threshold for co-methylation of 0.7 on the test set, is slightly better than AltumAge, with a mean square error of 25.28 compared to AltumAge's 28.31. 
    Overall, GraphAge performs slightly better than AltumAge. We analyzed error across age groups and sex (M/F). %\textbf{b.} In the case of MAE, we can see a mean absolute error of 3.20 compared to AltumAge's 3.29. Similar to MSE, we analyzed for different age groups and sex. 
    Detailed values for sub-figures a and b are in Table \ref{age_group_comp_table} in the Supplementary Materials. \textbf{c.} The bar plots compare age acceleration  for AltumAge and GraphAge models across various age groups in ovarian cancer, schizophrenia, and osteoporosis. For postmenopausal women with ovarian cancer, GraphAge generally predicts higher age acceleration compared to AltumAge, particularly in the 45-55 age group (3.452 vs. 0.670). This is consistent with the expectation of increased age acceleration in the presence of disease.In schizophrenia, the age acceleration values for both models are slightly negative, with AltumAge at -0.009 and GraphAge at -0.347, possibly due to the mental nature of the disorder. For osteoporosis, both models show negligible age acceleration, with GraphAge values closer to zero compared to AltumAge. In the 45-55 age group, age acceleration values are -6.786402 for AltumAge and -2.500751 for GraphAge. This aligns with existing knowledge, as no accelerated epigenetic aging is observed in osteoporotic patients' blood. \textbf{d.} We established that GraphAge's performance is competitive, with most predictions similar to AltumAge when compared to true chronological age. Even some age acceleration errors are similar. \textbf{e.} We can see that as we lower the threshold allowing the inclusion of more edges into the model, we decrease the test set error, suggesting that further improvement may be anticipated with a further decrease in the threshold.}

\label{performance_analysis}
\end{figure}

\subsection{Model Performance}
We trained GraphAge and, as has been mentioned above, compared it primarily with AltumAge~\cite{lapierre2022pan} for the reasons already mentioned above. We first split the data into a training and test sets (see Section \ref{sec:method}), then conducted a 5-fold cross-validation on the former, and subsequently, we tested the best model on the latter. Fig. \ref{performance_analysis} reports the model performances. GraphAge overall performance is quite competitive with AltumAge, showing in fact a slight improvement over the latter on the test set in our constrained compute setting (more on this later). In particular, the overall MAE (MSE) of GraphAge is \~3.207 (\~25.277) as opposed to \~3.296 (\~28.310) of AltumAge registering a modest improvement of 2.7\% (10.71\%) over the latter. Furthermore, we grouped all samples (in the test set) into age-specific groups and measured the model performance within each group. Fig. \ref{fig:mae} and Fig. \ref{fig:mse}, 
%(detail values along with number of samples can be found in Table \ref{age_group_comp_table} in the Supplementary Materials Section) 
illustrate the age error in different age (and sex) groups. We see similar but slightly higher values for MSE in female than male. The trend for MAE is similar across all age groups. 
%Starting from an MSE of 1.53 in male babies aged 0, it grows incrementally to 10.24 for ages 0-20 and 21.95 for ages 20-45. Then it decreases to 17.04 for ages 45-55, follows the trend to 28.12 for ages 55-65, rises sharply to 68.39 for ages 65-75, follows the trend to 38.40 for ages 75-80, and finally reaches 30.05 for males over 80. We see similar but slightly higher values for MSE in female as well. The trend for MAE is similar across all age groups. 
A consistent observation across all models is a decline in performance with increasing age. For example, starting from an MSE of 1.53 in male babies aged 0, it grows incrementally to 10.24 for ages 0-20 and 21.95 for ages 20-45. This supports the claim that epigenetic noise increases as chronological age diverges from epigenetic age \cite{lu2023information}. Fig. \ref{bio_vs_pred} demonstrates how both models predict epigenetic age in relation to true chronological age, showing that both are quite similar even in terms of age acceleration (i.e., the direction and value in which the predicted epigenetic age diverts from the chronological age.) on healthy datasets. 

Additionally, as an interesting case study, we tested both GraphAge and AltumAge on unhealthy datasets, i.e., in this experiment, the unhealthy samples were considered as unseen test samples for both the models. While this analysis also revealed similar age acceleration across all age groups for both the models (Fig. \ref{case study performance}), there are indeed some interesting points that are worth-discussing as follows. For postmenopausal women with ovarian cancer, GraphAge shows, on average, a (slightly) higher age acceleration, i.e., more in the positive direction than AltumAge. To elaborate, for age group 45-55, AltumAge shows an age acceleration value of 0.670 while GraphAge registers a much higher value of 3.452 and for age group 55-65, the values are -2.821574 and -0.783119 respectively. So, for this disease, GraphAge in most cases predicts a higher age acceleration than AltumAge, which is expected in case of a disease (in general). 

For the other two diseases, namely, schizophrenia and osteoporosis, a different phenomenon is noticed. In case of schizophrenia most samples are from age group 20-45. Here, the age acceleration values for both models are slightly negative: for AltumAge, it is -0.009141 and for GraphAge, -0.346602. This may be attributed to the fact that schizophrenia is a mental disorder and hence does not show much age acceleration. Finally for osteoporosis, for age group 45-55 (55-65), age acceleration values for AltumAge and GraphAge are -6.786402 (-6.978263) and -2.500751 (-3.728412), respectively. %For age 55-65 AltumAge shows -6.978263 and ours shows -3.728412. 
So, both models suggest a lack of age acceleration and GraphAge is more closer to zero. While this may sound surprising and counter-intuitive, this actually is inline with the current knowledgebase as no evidence for accelerated epigenetic age in blood of osteoporotic patients were found as reported in \cite{GSE99624}.

\subsection{Resource Constraints}
As has been alluded to above, we had to work in a resource constrained setting. To this end, Fig. \ref{threshold vs error} indicates that GraphAge performance is influenced by the threshold value (see Section \ref{sec:method}) used to filter edges from the underlying graph. Lowering the threshold value, which includes more edges in the underlying graph structure, enhances performance, albeit at the cost of increased computational cost. Just to provide an idea, the time for training one epoch increases from \~4 min to \~8 min when we decrease the threshold value from 0.8 to 0.7. Informatively, as we work in a constrained computational setting, we could not continue lowering the threshold value beyond the value reported here. However, Fig. \ref{threshold vs error} promises further performance improvement if more computational power could be made available. 

We further remark that, in parallel to performance improvement, the lowering of the threshold value also promises to provide more insightful interpretation as more edges are included in the underlying graph. As is evident from the Supplementary Materials (see Section \ref{Supplementary}), decreasing the threshold value from 0.75 to 0.70 increases the edges by 90\% thereby potentially making the graph more informative. %As shown in Supplementary Materials we see that decrease in threshold from 0.75 to 0.70 has seen an increase of 90\% edge number increase.

\subsection{Interpretation}

%\subsubsection{Node Attributes (i.e., CpG Site Information) Analysis}
\subsubsection{Node Attributes Analysis}
\begin{figure}[htbp!]
    \centering
    \begin{subfigure}[b]{0.4\textwidth}
        \centering
        \includegraphics[width=\textwidth]{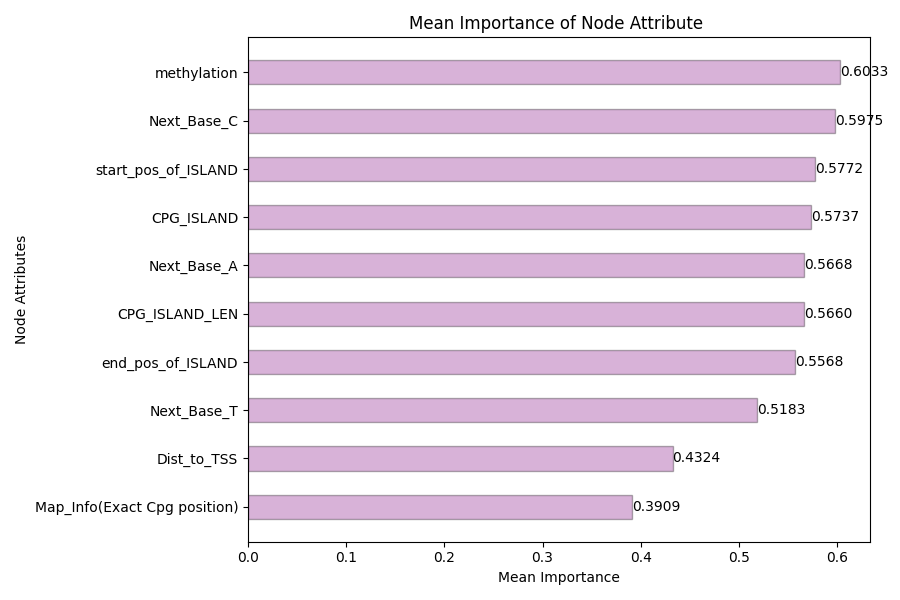}
        \caption{}
        \label{node mean importance}
    \end{subfigure}
    \begin{subfigure}[b]{0.4\textwidth}
        \centering
        \includegraphics[width=\textwidth]{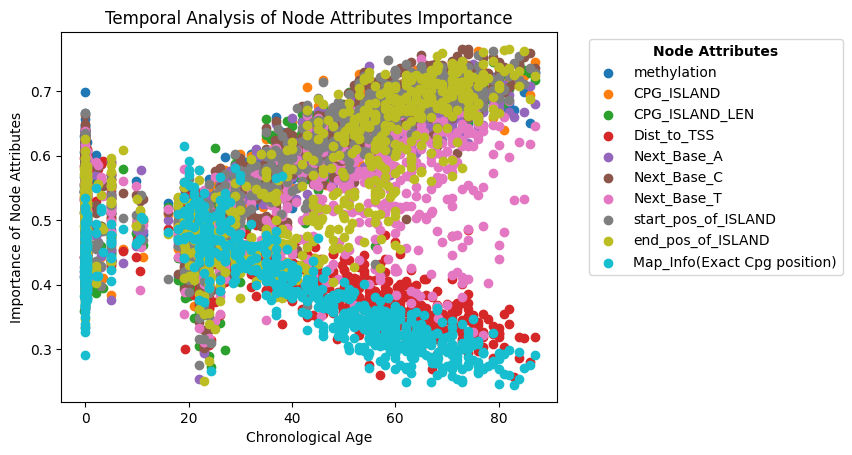}+
        % {images/node_attr_temp.png}
        \caption{ }
        \label{temporal analysis of node importance}
    \end{subfigure}
    \caption{\textbf{a.} Using the GNN explainer, we determined the importance of each node attribute (please refer to Table \ref{Table: Node attribute variable des} of Supplementary Materials for a description thereof) for every individual sample. We saved these results in a dataset and analyzed them in a new notebook. The importances were then averaged to provide a broader understanding of which attributes are most influential. We observe that the methylation beta value is the most important factor in predicting age. Following this, we see some interesting observation. We see that the information about CpG islands and the next base pairs is very important for predicting age. We can see that whether the next base pair is C is quite significant. This is followed by the start position of the island and whether the node (i.e., CpG site) is a CpG Island. Additionally, the length of the island is also prominent, indicating that being a CpG Island has a significant effect on the aging process. We can also see the effect of other next base pairs in the CpG site. Conversely, the distance to the transcription start site (Dist\_TSS) and the average position of the site are less influential. Therefore, the location of a CpG Island and its length, along with the starting position of the island and the next base pair, play a crucial role in the aging process.\textbf{b.} %We also performed a temporal analysis by plotting the importance with respect to chronological age. We observe that the distance to the transcription start site (Dist\_TSS) and MapInfo (the average position of the site) see a slight increase up to the 20s, whereas the rest decrease up to the 20s. Afterwards, they do the opposite: Dist\_TSS and MapInfo (average position) decrease in importance, while the rest increase in importance. 
    In our temporal analysis (importance score vs, chronological age), we observe that Dist\_TSS and Map\_Info see a slight increase up to the 20s, whereas the rest decrease up to the 20s. Afterwards, they do the opposite: Dist\_TSS and Map\_Info decrease in importance, while the rest increase.
}

    \label{node analysis}
\end{figure}

We identified important nodes, edges, and node attributes for each sample. Recall that, node attributes actually correspond to CpG Site information. We analyzed these important factors in detail as follows. First, we averaged the importance scores of node attributes across all samples to demonstrate our model’s positional awareness through its feature importance (Fig. \ref{node mean importance}). From the mean importance, we can see that methylation is highly significant as expected. Additionally, the next base pair of all CpG sites, the characteristics of being in a CpG island (CPG\_ISLAND) and the starting and ending position of the island (start\_pos\_of\_ISLAND, end\_pos\_of\_ISLAND) along with its length (CPG\_ISLAND\_LEN) are crucial factors according to our analysis. The association of CpG island with aging has also been found in various studies in the literature (e.g., \cite{issa2000cpg,toyota1999cpg,christensen2009aging}) and variations in the length thereof are shown to influence gene regulation complexity and evolutionary mechanisms \cite{elango2011functional}, which in turn can influence aging \cite{perez2020aging,holzscheck2021modeling}. Relevantly, CpG islands are often located near the promoters of genes so their starting and ending positions are more informative \cite{illingworth2009cpg}. 

Our analysis also reveals that the base pair adjacent to a CpG site (Next\_Base\_C, Next\_Base\_A and Next\_Base\_T) also plays a significant role in the aging process. This may be attributed to the fact that the flanking sequence (the sequence adjacent to a CpG site) can influence its likelihood of being methylated, suggesting that the surrounding sequence directly impacts methylation dynamics \cite{santoni2021impact} and hence, aging. We additionally found that Cytosine (C) next to a CpG site (Next\_Base\_C) has a greater influence on aging than Adenine (A) or Thymine (T) (Next\_Base\_A, Next\_Base\_T). This seems interesting as Gao et. al. \cite{gao2020comprehensive} recently showed that the enzyme DNMT3A (DNMT3B) prefers to methylate CpG sites followed by Cytosine or Thymine (Guanine or Adenine), with Cytosine (Guanine) being more influential than Thymine (Adenine). So, GraphAge was able to capture the importance of Cytosine next to a CpG site accurately. Informatively, the results reported in Fig. \ref{node mean importance} is for threshold value 0.70. We also provided results with threshold value 0.75 in the Supplementary Materials Section (Fig. \ref{node_mean_importance_0.75}).

\subsubsection{Methylation Regulated Networks and subnetworks}
\begin{figure}[htbp!]
    \centering
        \begin{subfigure}[b]{0.8\textwidth}
            \centering
            \includegraphics[width=\textwidth]{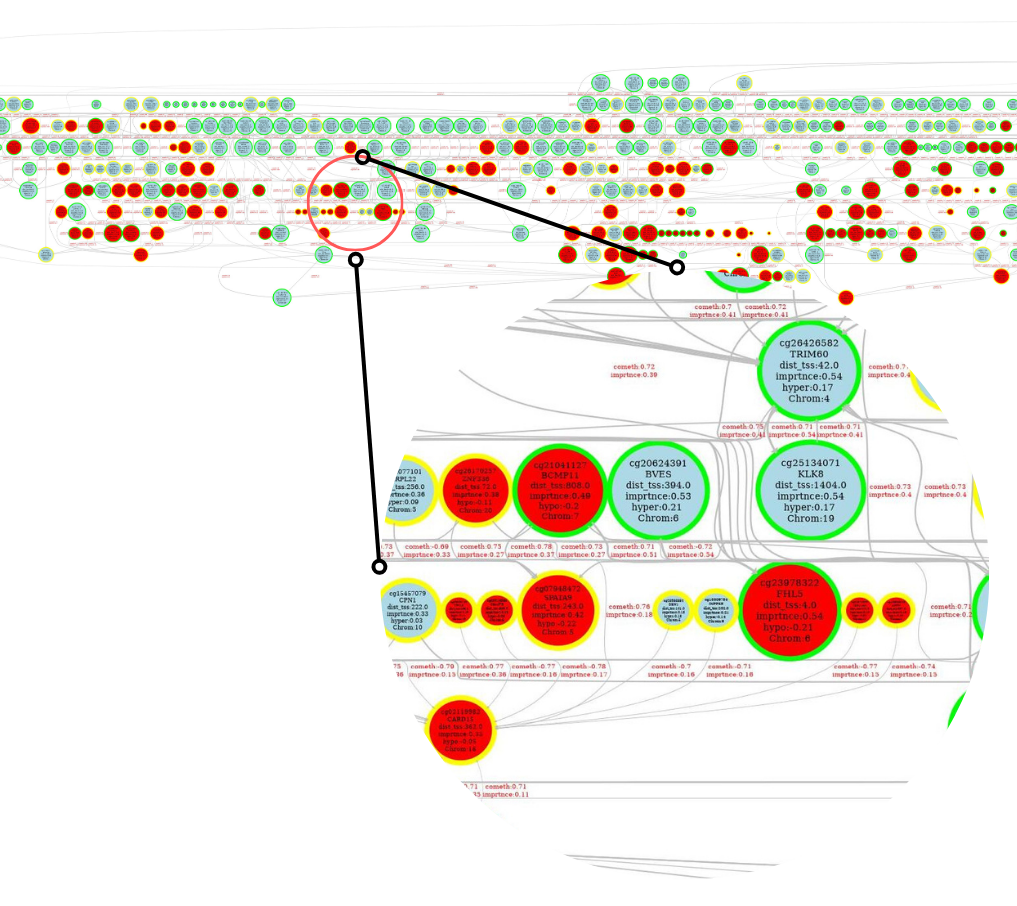}
            \caption{}
            \label{subnetwork graph}
        \end{subfigure}
        \begin{subfigure}[b]{0.4\textwidth}
            \centering
            \begin{subfigure}[b]{0.4\textwidth}
                \centering
                \includegraphics[width=\textwidth]{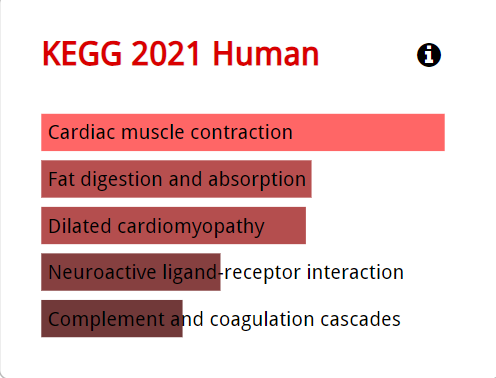}
                \label{kegg hyper}
            \end{subfigure}
            \begin{subfigure}[b]{0.4\textwidth}
                \centering
                \includegraphics[width=\textwidth]{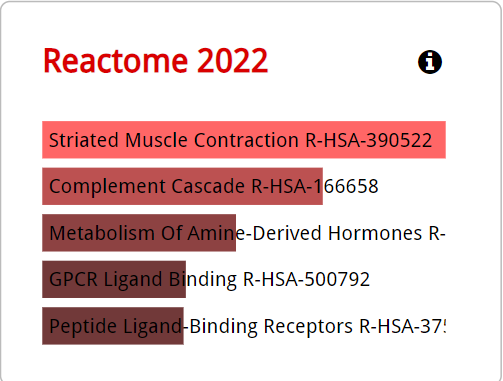}
                \label{reactom hyper}
            \end{subfigure}
            \caption{}
            \label{Hyper explain}
        \end{subfigure}
        \begin{subfigure}[b]{0.4\textwidth}
            \centering
            \begin{subfigure}[b]{0.4\textwidth}
                \centering
                \includegraphics[width=\textwidth]{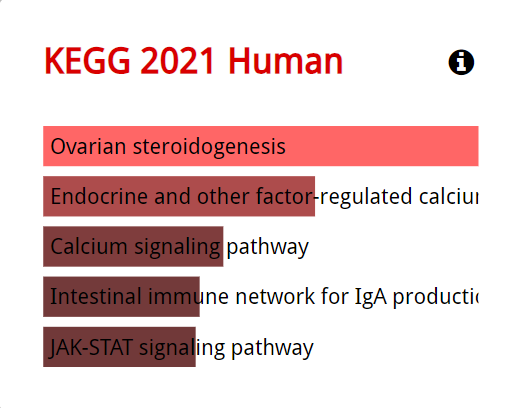}
                \label{kegg hypo}
            \end{subfigure}
            \begin{subfigure}[b]{0.4\textwidth}
                \centering
                \includegraphics[width=\textwidth]{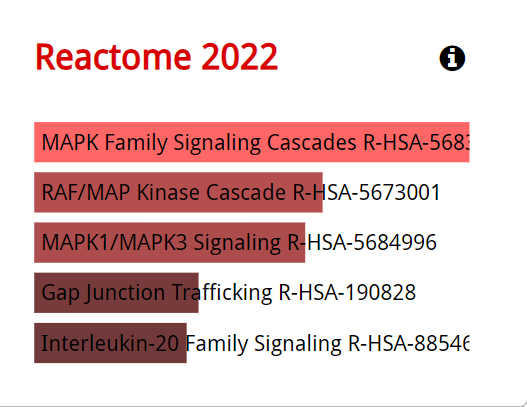}
                \label{reactom hypo}
            \end{subfigure}
            \caption{}
            \label{Hypo explain}
        \end{subfigure}
        \caption{
\textbf{a.}GNN explainer outputs node and edge importance for each sample. We averaged these values across the same age groups used to show test set errors. After averaging, we excluded zero-importance nodes and low-importance edges, forming subnetworks of connected nodes. Here, we show one such subnetwork and a magnified version, annotated with importance and methylation status. Red nodes are hypomethylating, blue nodes are hypermethylating, green circles indicate increasing importance, and yellow circles indicate decreasing importance. See Section \ref{Color code} for details. Additionally, we identified the genes of the CpG sites in the subnetwork. Based on whether they are hypo- or hypermethylating with age, we separated the genes and performed Enrichr analysis. \textbf{b.} Here, we show the pathway analysis for the genes that become hypermethylating with age. The analysis indicates that the Cardiac Muscle contraction pathway becomes hypermethylating with age, leading to reduced gene expression.  \textbf{c.} We also identified the pathway of genes that become hypomethylating with age as Ovarian steroidogenesis pathway becomes hypomethylating, resulting in increased gene expression. These pathways in b and c also affect each other in the aging process which can be seen in \textit{a}. 
}

        \label{subnetwork}
\end{figure}

We determined the importance of nodes and edges for each sample data point. We then divided the samples into age groups (the same age groups used during testing) and averaged the importance of the samples in each age group. We eliminated all nodes having zero importance and all low-importance edges (i.e., having scores less than 0.1). Following this procedure we identified subnetworks of important (connected) CpG sites, which we termed as Methylation Regulated Networks (MRN). The rationale behind this name is that methylation regulates gene expression, and these important sites work together through co-methylation, forming a network that contributes significantly to the aging process. We generated these graphs for all age groups and identified all subnetworks within the MRN 
%created separate images of all the subnetworks 
that contain more than a specific number (in particular, 10) of CpG sites. Fig. \ref{subnetwork graph} shows a single segment of our entire graph from age group 20-45 representing a MRN. All the MRNs and subnetworks in the MRN with more than 10 CpG\_sites (segments) of each age group can be found at \cite{bojack2024graphage}.  
%subnetwork of connected CpG sites. 
%For the ease of researchers the graphs are color-coded based on the rules of Section \ref{Color code} and annotated with various information explained in \ref{graph annotation}. 
These MRNs (along with identified subnetworks therein) provide us with a clearer view to visually see the relationships among CpG sites for further meaningful analysis. A number of such analyses (non-exhaustive though) are reported in the rest of this section.
%as can be seen from Fig. \ref{subnetwork graph} and Fig. \ref{fig:directionality and colour code}. 
%%%%%%%%%%%%%%%%%%%%%%%%%%%%%%%%%%%%%%%%%% SUBNETWORK IMAGE%%%%%%%%%%%%%%%%%%%%%%%%%%%%%

%%%%%%%%%%%%%%%%%%%%%%%%%%%%%%%%%%%%%%

\begin{figure}
    \centering
    \includegraphics[width=\linewidth]{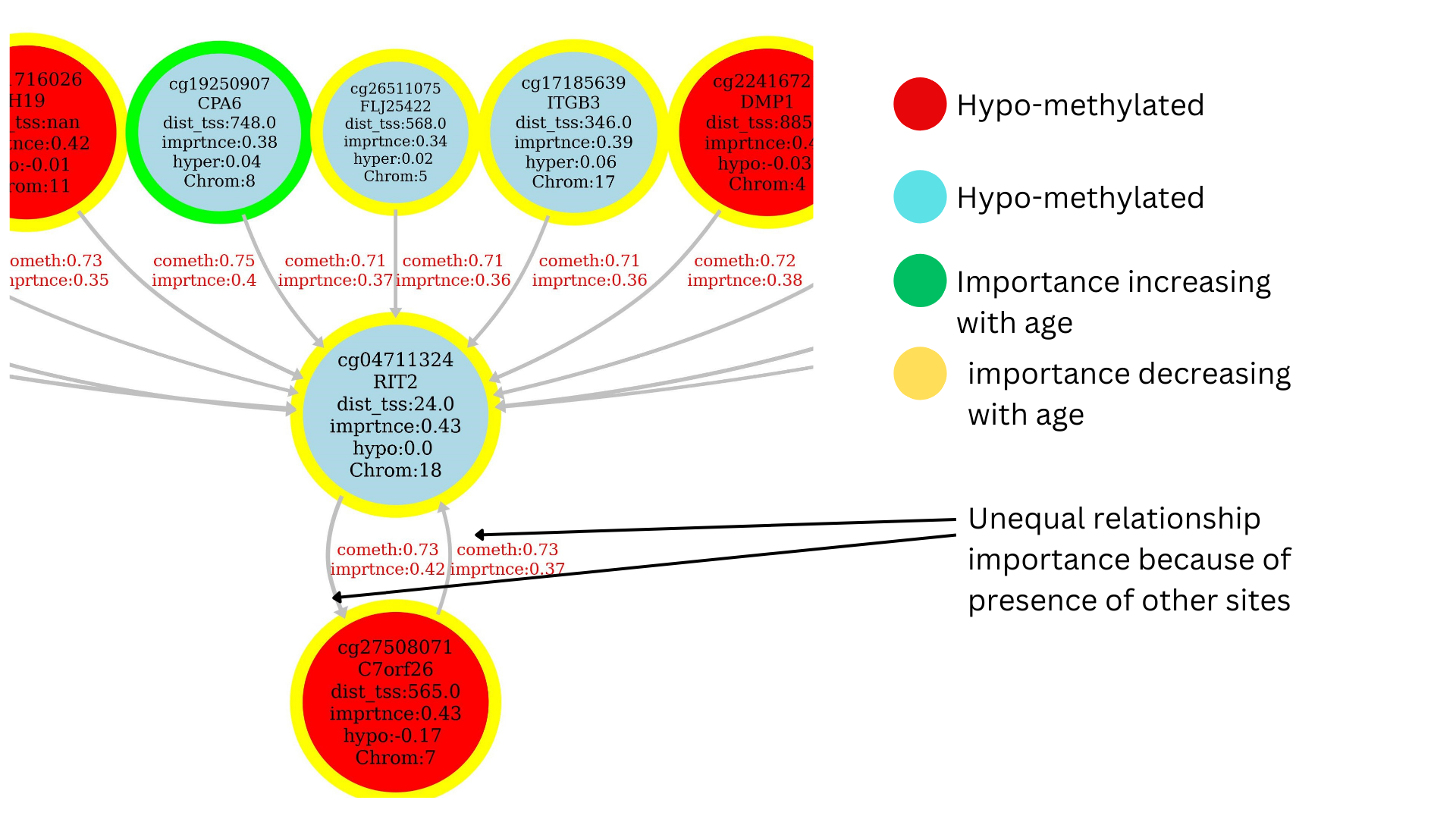}
    \caption{The methylation event from site cg04711324 to site cg27508071 might be important as it is influenced by subsequent changes in sites cg19250907, cg26511075 and others, creating a chain reaction. However, the reverse (cg27508071 to cg04711324) does not necessarily hold the same importance. This directional influence underlines the complexity and interconnected nature of methylation regulation within the network}
    \label{fig:directionality and colour code}
\end{figure}

The directionality of the MRNs provides interesting insights (Fig. \ref{fig:directionality and colour code}). For example, the methylation event from site cg04711324 to site cg27508071 seems to be important as it is influenced by subsequent changes in sites cg19250907, cg26511075 and others, creating a chain reaction. However, the reverse (cg04711324 to cg27508071) does not necessarily hold the same importance. This directional influence underlines the complexity and interconnected nature of methylation regulation within the network. %Such a scenario is shown in Fig. \ref{fig:directionality and colour code}. 
Notably, other models in the literature, including AltumAge, are unable to capture this level of complexity and interconnected relationships.%, which becomes evident through such detailed visualizations and subsequent analyses.

\subsubsection{Pathway analyses}
\label{sec: pathway analysis}
We separated the genes of each MRN for each age group based on whether the nodes are hypo- or hypermethylating with the progression of age. We then performed pathway analysis using Enrichr \cite{chen2013enrichr, kuleshov2016enrichr, xie2021gene} for these two types of genes separately. Fig. \ref{Hyper explain} and Fig. \ref{Hypo explain} present the pathway analysis of the hypermethylating and hypomethylating genes, respectively, from the subnetwork in Fig. \ref{subnetwork graph}. From Fig. \ref{Hyper explain}, 
%which shows the pathway analysis of hypermethylating genes in the previously mentioned subnetwork graph, 
we can see that the genes under consideration belong to the Cardiac Muscle Contraction pathway, which is crucial for aging and becomes hypermethylating with the progression of age, thereby decreasing gene expression. This is interesting as vascular disease is influenced by hypermethylation \cite{xu2021roles, boovarahan2022age} and since cardiac muscle is known to change with aging \cite{lakatta1987cardiac}. 

A similar analysis reveals that the hypomethylating genes under consideration belong to the Ovarian Steroidogenesis pathway (Fig. \ref{Hypo explain}) thereby identifying it as a significant pathway in aging, which, with age, gets hypomethylating resulting in increased gene expression. This seems interesting as a recent study focused on children associated hypomethylation in several genes (e.g., CYP11A1, CYP19A1) with steroidogenesis \cite{vanhorebeek2023abnormal}. An earlier study also associated aging-related DNA methylation (dominantly, hypomethylation) patterns in peripheral blood with ovarian cancer \cite{teschendorff2009epigenetic}.

At this point a brief discussion on a unique analytical strength of GraphAge is in order. Other models like AltumAge can also conduct pathway analyses, albeit with an important distinction in respect to GraphAge as follows. In AltumAge, the pathways are selected based only on the `important' CpG sites identified through some feature selection mechanism (e.g., SHAP values), without any consideration regarding relations/interactions among them. Unlike its competitors, GraphAge on the other hand has the capability to consider such interactions/relations in addition to considering the importance score of the CpG sites, which has been demonstrated here.

\subsubsection{Analyzing aging-related pathways}
We investigated three aging-related pathways, namely, SIRT, mTOR, and AMPK \cite{zhao2020sirtuins,weichhart2018mtor,burkewitz2014ampk}, which were also investigated in \cite{lapierre2022pan} while interpreting AltumAge. Surprisingly, AltumAge could not establish any particular relevance with respect to two of them, namely, mTOR and SIRT. %, with its most important CpG site being cg04508649 (SHAP importance =0.0062, rank 7705). 
GraphAge, however, was able to identify highly contributing genes from the mentioned pathways. Below we briefly report the findings.

\begin{description}
    \item[\textbf{mTOR pathway.}] For this analysis, We got the gene set for mTOR pathway from \cite{Harmonizome_mTOR_Signaling_Pathway}. GraphAge identified a number of its CpG sites (e.g., CpG sites located in genes TSC1 and PDPK1) as important in different age groups. In particular, CpG site cg14444710, located in gene PDPK1, was found to have the highest importance score for Age group 0 (importance score: 0.528059; 98.70th percentile), Age group 0-20 (importance score: 0.503974; 97.98th percentile), and Age group 20-45 (importance score: 0.451429; 91.79th percentile). 

    Similarly, CpG site cg19393006, located in gene TSC1, was found to have the highest importance score for Age group 45-55 (importance score: 0.515012; 94.15th percentile), Age group 55-65 (importance score: 0.565572; 91.49th percentile), Age group 65-75 (importance score: 0.569637; 90.44th percentile), Age group 75-80 (importance score: 0.588715; 86.23rd percentile), and Age group 80+ (importance score: 0.60006; 86.11th percentile).  
    
    \item[\textbf{SIRT pathway.}] For SIRT, we identified the corresponding gene set from \cite{qiagen_sirtuin_pathway}. GraphAge identified a number of its CpG sites (e.g., CpG sites located in MAPK4 and RELA as the most dominating ones in this context) as important in different age groups. We found CpG site cg26946769, located in gene MAPK4, to have the highest importance score in Age group 0 (importance score: 0.484208; 96.40th percentile) and Age group 45-55 (importance score: 0.618472; 98.07th percentile). On the other hand, CpG site cg18746357, located in RELA, was found to have the highest importance score for Age group 0-20 (importance score: 0.470072; 93.85th percentile), Age group 20-45 (importance score: 0.51333; 96.98th percentile), Age group 55-65 (importance score: 0.671955; 99.27th percentile), Age group 65-75 (importance score: 0.688102; 99.02th percentile), Age group 75-80 (importance score: 0.688684; 97.08th percentile), and Age group 80+ (importance score: 0.698615; 97.92th percentile). 
    % GraphAge identified a number of its CpG sites (e.g., CpG sites located in MAPK4 and RELA as the most dominating ones in this context) as important in different age groups. We found CpG site cg26946769, located in gene MAPK4, to have the highest importance score in Age group 0 (importance score: 0.484208) and Age group 45-55 (importance score: 0.618472). On the other hand, CpG site cg18746357, located in RELA, was found to have the highest importance score for Age group 0-20 (importance score: 0.470072), Age group 20-45 (importance score: 0.51333), Age group 55-65 (importance score: 0.671955), Age group 65-75 (importance score: 0.688102), Age group 75-80 (importance score: 0.688684) and Age group 80+ (importance score: 0.698615). 

    \item[\textbf{AMPK pathway.}] Recall that, both the models found the third pathway, namely, AMPK, relevant and important, albeit with different explanation. AltumAge identified CpG site cg22461835, located in ADRA1A, as important (through SHAP-based analysis). However, GraphAge differs with this finding as follows. We collected the gene set for AMPK from \cite{qiagen_ampk} and in our analysis,   
    %as CpG site cg22461835, located in ADRA1A, was not found to be significant (importance score :0) in aging process. Rather, 
    GraphAge identified other genes to be of significant relevance, e.g., GNAS, TBC1D1, LEP. In particular, CpG site cg25268451, located in gene GNAS, was found to have the highest value in importance for Age group 0 (importance score: 0.524305; 98.45th percentile) and CpG site cg25608041, located in TBC1D1, stood out in Age group 0-20 (importance score: 0.520926; 99.44th percentile). On the other hand, CpG site cg19594666, located in gene LEP, was found to have the highest value in importance for Age group 20-45 (importance score: 0.498691; 96.04th percentile), Age group 45-55 (importance score: 0.615761; 97.85th percentile), Age group 55-65 (importance score: 0.634677; 95.24th percentile), Age group 65-75 (importance score: 0.638405; 93.71th percentile), Age group 75-80 (importance score: 0.671997; 92.46th percentile), and Age group 80+ (importance score: 0.675434; 92.77th percentile).
    % GraphAge identified other genes to be of significant relevance, e.g., GNAS, TBC1D1, LEP. In particular, CpG site cg25268451, located in gene GNAS, was found to be have the highest value in importance for Age group 0 (importance score: 0.524305) and CpG site cg25608041, located in TBC1D1, stood out in Age group 0-20 (importance score: 0.520926). On the other and, CpG site cg19594666, located in gene LEP, was found to be having the highest value in importance for Age group 20-45 (importance score: 0.498691), Age group 45-55 (importance score: 0.615761), Age group 55-65 (importance score: 0.634677), Age group 65-75 (importance score: 0.638405), Age group 75-80 (importance score: 0.671997), and Age group 80+ (importance score: 0.675434).
    \item[\textbf{Genes in these pathways.}] Before concluding this section, we briefly remark on some dominating genes (according to GraphAge interpretation) from these three pathways regarding their association with respect to aging. We observed in our analysis that TSC1 and PDPK1 were repeated more times than others for mTOR pathway, while RELA and LEP were repeated more times for the other two pathways. Now, TSC1 upregulation in senescent fibroblasts is associated with aging-related changes \cite{fellows2012age}. PDPK1 has been found highly relevant for 
    human longevity \cite{serbezov2019pool}. On the other hand, in adult humans of different body weight, serum leptin (LEP) gradually declines during aging \cite{isidori2000leptin}. Finally, expression of NF-KB subunits (such as RELA) has been correlated with age in \cite{helenius1996aging}. Also, we remark that, during the growth stage from 0 to 20 years, BC1D1, the most dominant one among the AMPK pathway genes (according to our analysis), plays a role in regulating cell growth and differentiation \cite{ncbi_lep}. Further to above discussion, there are multiple studies on mice that linked these genes with aging (please refer to Section \ref{sec:age_related_further_analysis} of Supplementary Section for a brief discussion). 
    
    %For example, TSC1, which is a tumor suppressor gene \cite{ncbi_TSC1}, is shown to have association with accelerated retinal aging in knockout mice \cite{rao2021mtorc1} and moderate lifelong overexpression thereof has been shown to improve health and survival in mice \cite{zhang2017moderate}. On the other hand, aging affects leptin (LEP) actions differently, with a decline in hypermetabolic responses but increased sensitivity in lean rats. Old rats are more prone to obesity-induced leptin resistance, while calorie restriction enhances leptin responsiveness, particularly in older rats \cite{balasko2014leptin}.  

% Studies have shown the relationship between age and carcinogenesis involves the accumulation of somatic mutations that activate oncogenes and inactivate tumor suppressor genes (such as TSC1) \cite{fearon1990genetic}. Genes like MAPK4, BCL2, RelA, LEP, and GNAS contribute to the development and progression of carcinogenesis \cite{wang2022mapk4,ncbi_TSC1,nci_bcl2,sethi2008nuclear,hosoda2015gnas,caruso2023leptin}. 
%Additionally, during the growth stage from 0 to 20 years, BC1D1, presumed to play a role in regulating cell growth and differentiation \cite{ncbi_lep}, was the most dominant among the AMPK pathway genes.
\end{description}

\subsubsection{Temporal Analysis}

We conducted temporal analyses along multiple dimensions and tried to unearth some important insights. %, namely on node attributes and on CpG sites and associated genes. 
Below we briefly present our findings.

\begin{figure}[htbp!]
    \centering
    \begin{subfigure}[b]{\textwidth}
        \centering
            \begin{subfigure}[b]{0.35\textwidth}
            \centering
            \includegraphics[width=\textwidth]{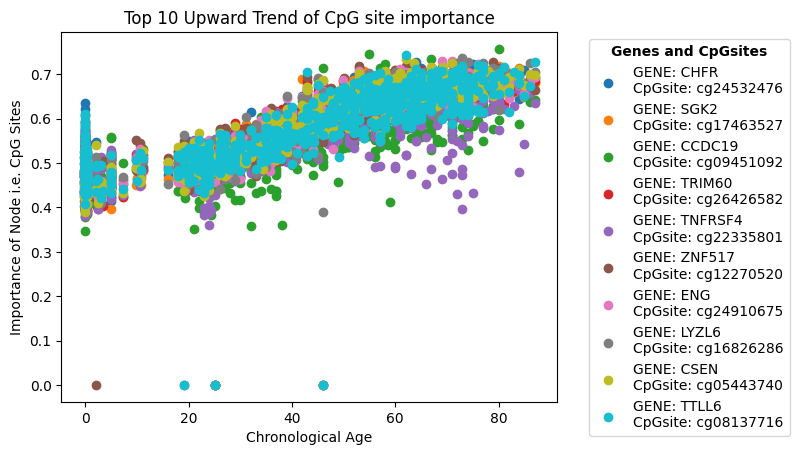}
            \caption{}
            \label{upward}
        \end{subfigure}
        \begin{subfigure}[b]{0.35\textwidth}
            \centering
            \includegraphics[width=\textwidth]{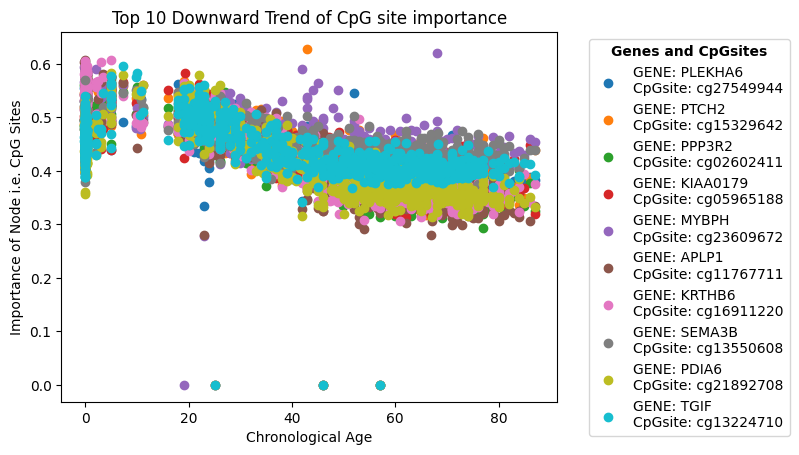}
            \caption{}
            \label{downward}
        \end{subfigure}
        \begin{subfigure}[b]{0.35\textwidth}
            \centering
            \includegraphics[width=\textwidth]{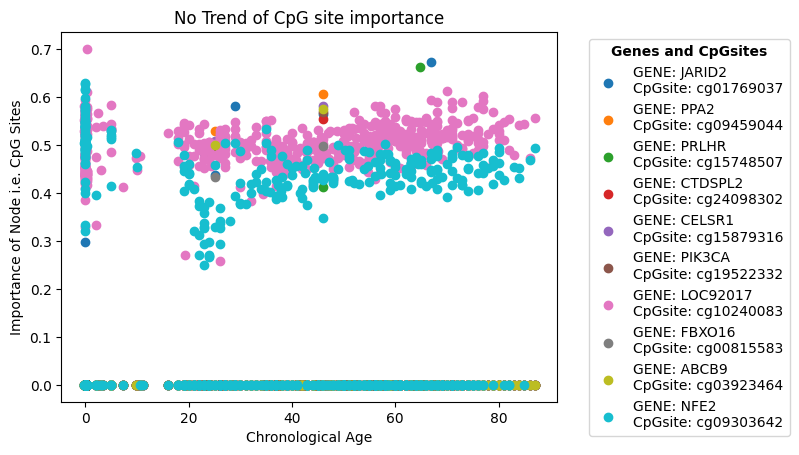}
            \caption{}
            \label{no_trend}
        \end{subfigure}
        \caption*{}
        \label{}
    \end{subfigure}
    \begin{subfigure}[b]{0.5\textwidth}
    \centering
        \begin{subfigure}[b]{0.35\textwidth}
            \centering
            \includegraphics[width=\textwidth]{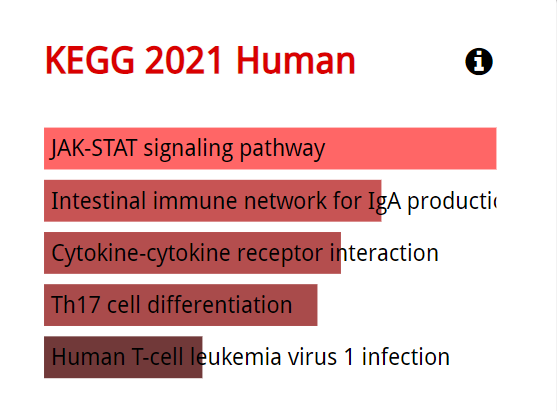}
            \caption{}
            \label{inc_hypo}
        \end{subfigure}
        \begin{subfigure}[b]{0.35\textwidth}
            \centering
            \includegraphics[width=\textwidth]{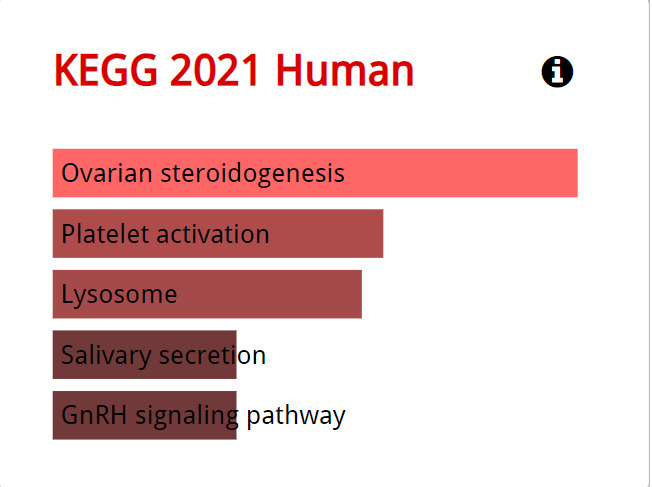}
            \caption{}
            \label{dec_hypo}
        \end{subfigure}
    \end{subfigure}
    \begin{subfigure}[b]{0.5\textwidth}
    \centering
            \begin{subfigure}[b]{0.35\textwidth}
                \centering
                \includegraphics[width=\textwidth]{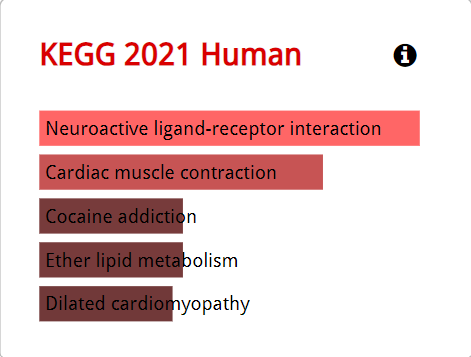}
                \caption{}
                \label{inc_hyper}
            \end{subfigure}
            \begin{subfigure}[b]{0.35\textwidth}
                \centering
                \includegraphics[width=\textwidth]{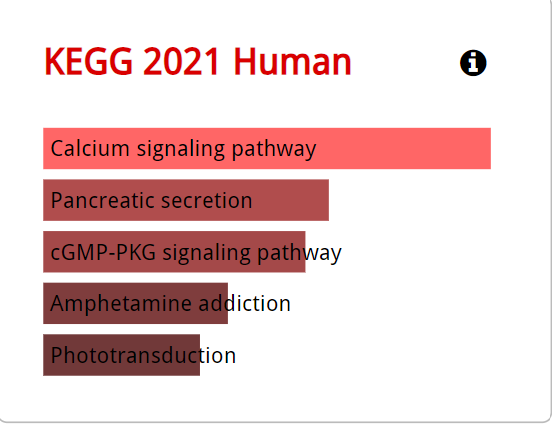}
                \caption{}
                \label{dec_hyper}
            \end{subfigure}
    \end{subfigure}

   \caption{
GNN explainer assigns importance to nodes and edges for each sample, allowing us to observe how the importance of these nodes, i.e., CpG sites, varies with age. We used a linear regression model to find the slope of each importance with respect to age. \textbf{a.} Here we show the top 10 upward-trending CpG sites in terms of importance and their associated genes. \textbf{b.} Here we show the top 10 downward-trending CpG sites in terms of importance and their associated genes. \textbf{c.} Some CpG sites' importance does not vary with age. Here we show those sites and their associated genes. \textbf{d.} Based on these values, we color-coded our MRN (Methylation Regulated Network) with green circles (importance increasing) and yellow circles (importance decreasing) in Fig. \ref{subnetwork graph}. We categorized hypomethylating genes based on whether CpG site importance scores are increasing or decreasing. Our pathway analysis revealed that the importance of the JAK-STAT pathway is increasing. \textbf{e.} Conversely, pathway analysis on genes with decreasing importance showed that the importance of the ovarian steroidogenesis pathway is decreasing. \textbf{f.} We repeated the above analysis for hypermethylating genes and found that the importance of the Neuroactive Ligand-Receptor Interaction pathway is increasing. \textbf{g.} Additionally, by performing pathway analysis on the importance-decreasing genes, we discovered that the importance of the calcium signaling pathway is decreasing.
}

    \label{Node importance temporal analysis}
\end{figure}

\begin{description}
    \item[\textbf{Analysis on Node Attributes.}] We examined the importance of each node attribute throughout the aging process (Fig. \ref{temporal analysis of node importance}). We observe that the CpG related information (CPG\_ISLAND, CPG\_ISLAND\_LEN, start\_pos\_of\_ISLAND and end\_pos\_of\_ISLAND) become increasingly important, particularly age 30 onward. This pattern is in line with the findings of different studies (e.g., \cite{issa2000cpg,toyota1999cpg,christensen2009aging}) in the literature that were carried out on `young' subjects (e.g., age range 30 and above). A closer look in Fig. \ref{temporal analysis of node importance} in the region before age 30, reveals further interesting insight. During birth and infancy, various information about CpG Island also shows high importance. But subsequently, this changes, i.e., importance score slowly decreases, till the age is around 30. So, this age range of 5-30 could be investigated further, particularly from a biological point of view. Informatively, we notice the same pattern for flanking sequence, i.e., the sequence adjacent to a CpG site (Next\_Base\_A, Next\_Base\_T). Informatively, the results reported in Fig. \ref{temporal analysis of node importance} is for threshold value 0.70. We also provided results with threshold value 0.75 in the Supplementary Materials Section (Fig. \ref{temporal_analysis_of_node_importance_0.75}).

    % As mentioned before, this analysis is for threshold 0.70 in Fig. \ref{temporal analysis of node importance} but we had conducted analysis for 0.75 in Fig. \ref{temporal_analysis_of_node_importance_0.75} in the Supplementary Materials Section. The analysis was found to be consistent with the one in threshold 0.70.
    
    \item[\textbf{Analysis on CpG sites and associated genes.}] We conducted another temporal analysis on the importance of CpG sites (i.e., the nodes) and their associated genes, demonstrating how the role of each CpG site varies with age. Fig. \ref{upward} shows the top 10 CpG sites and their associated genes that exhibit an upward trend in importance, while Fig. \ref{downward} depicts the top 10 sites with a downward trend. As shown in Fig. \ref{no_trend}, some CpG sites' roles do not vary significantly with age. 
    
    \item[\textbf{Analyzing hypomethylating genes.}] We categorized the previously mentioned hypomethylating genes into two groups based on whether the importance score of the corresponding CpG sites are increasing or decreasing according to our temporal analysis and found some interesting insight. 
    For example, our pathway analysis revealed that the importance of the JAK-STAT pathway is increasing (Fig. \ref{inc_hypo}) and that of the ovarian steroidogensis pathway is decreasing (Fig. \ref{dec_hypo}). Now, increasing JAK-STAT signaling in adulthood is known to increase levels of inflammatory cells in aging muscle tissue \cite{price2014inhibition} and there are studies that have linked these two pathways together (e.g., \cite{zareifard2023janus,wang2023role}). Also, recall that the ovarian steroidogenesis pathway was found as a significant pathway with respect to aging in our earlier analysis (Section \ref{sec: pathway analysis}). Thus our finding that for these two linked pathways that while one becomes more important the other becomes less, could be an interesting addition to the current knowledge base, which of course necessitates further biological validation.
    %This also corresponds with studies which have linked these two together. Zareifard et. al. show that JAK-STAT pathway plays a crucial role in ovarian steroidogenesis by mediating the effects of hormones such as FSH and LH. When these hormones bind to their receptors on ovarian cells, they activate the JAK kinases, which in turn phosphorylate STAT proteins. The activated STATs translocate to the nucleus and promote the transcription of genes essential for steroid hormone biosynthesis, such as aromatase and StAR. This pathway ensures the proper production of estrogen and progesterone, crucial for ovarian function and overall reproductive health \cite{zareifard2023janus}. Another study also found that, the activation of the p-JAK2/p-STAT3 signaling pathway has been implicated in follicular development in PCOS rats. This pathway may influence follicular maturation by affecting the expression of candidate genes such as LHCGR, FSHR, CYP17a, and CYP19, which are crucial for the regulation of gonadotropin receptors and enzymes involved in steroidogenesis \cite{wang2023role}. 
    %So these studies agree that they both have an influence and our results show that as one's importance increases with age, other's one goes down. 
    \item[\textbf{Analyzing hypermethylating genes.}] We also repeat the above analysis for hypermethylating genes.% (which were identified as cardiac muscle contraction found in the Section \ref{sec: pathway analysis}). 
    In this case, for example, We found that the importance of Neuroactive Ligand-Receptor Interaction pathway is increasing (Fig. \ref{inc_hyper}) while that of the calcium signaling pathway is decreasing (Fig. \ref{dec_hyper}). Both of these pathways are shown to have association with aging (e.g., \cite{ge2023retinol,berridge2016vitamin}) and there are evidences in the literature regarding links between the two (e.g., \cite{zheng2007calcium,fang2021enrichment}).   
    
    %Neuroactive Ligand-Receptor Interactions are key pathways involved in the aging of skeletal muscle stem cells \cite{ge2023retinol}. And research also show role of calcium signalling in ageing \cite{berridge2016vitamin}. Finally we also realize that they are related. This corresponds with study showing a connection between neuroactive ligand and calcium signalling pathways \cite{zheng2007calcium}. A study also finds that signaling pathways such as neuroactive ligand-receptor interaction, the calcium ion signaling pathway, and cardiac muscle contraction may be related to the pathogenesis of CHF (Chronic Heart Failure) and how they are related \cite{fang2021enrichment}.    

\end{description}

\subsubsection{Visualizing temporal aging signal}
Before concluding this section a brief discussion on the visualization of the temporal signals is in order. We have integrated this ``temporal aging signal" into our MRN images (e.g., Fig. \ref{subnetwork graph} and Fig. \ref{fig:directionality and colour code}) using color coding: green circles are used for increasing importance and yellow for decreasing importance. This visualization provides researchers with an intuitive understanding of how sites interact and influence each other's importance. 
As illustrated in Fig. \ref{subnetwork graph}, many yellow-circled (decreasing importance) sites are directly connected to green-circled (increasing importance) sites, along with other annotations that help elucidate the complex relationships among different sites. While AltumAge also attempted a similar analysis by introducing interaction values (leveraging SHAP importance) to understand interactions, GraphAge, with its structural information provides a broader and more comprehensive view of these relationships through MRNs.

\FloatBarrier

%\section{Discussions}\label{Discussion}

\section{Conclusion}\label{Conclusion}
GraphAge has been able to predict age by utilizing all available information and leveraging the GNN explainer, we gained various insights into how CpG sites interact and how their importance change with the progression of age through the temporal analysis. We are able to identify pathways significant for aging. We also analysed the importance of some well known aging-related pathway gene in our model's analysis. Moreover We observe how the importance of node attributes, such as, various biological information of CpG sites, influences the aging process. We also noted how their importance varies with age. 
%From our research, we found that CpG islands have a significant impact on the aging process. We also discovered that the nucleotide next to the CpG site contributes notably to the aging process. Moreover, we found that Cytosine (C) next to a CpG site has a greater influence on methylation than Adenine (A) or Thymine (T). 
% Though this particular comparison is not seen but we can understand that this C is not in a CpG site and DNA methylation is addition of methyl group onto the C5 position of the cytosine to form 5-methylcytosine and the majority occurs in CpG site  \cite{moore2013dna,jabbari2004cytosine}.But some study show that non-CpG sites (such as CpT, CpA or even CpC) can also be methylating \cite{jang2017cpg}. And increasing evidence suggests that non-CpG methylation, particularly abundant in the adult human brain, plays a crucial role in gene regulation and is influenced by DNMT3A and DNMT3B, impacting aging and age-associated diseases \cite{pinney2014mammalian}. And studies have found that methylation of non-CpG sites contribute to various disease, such as, Cancer \cite{li2019hypomethylation}. Another study show  non-CpG methylation may play a role in the pathophysiology of SCI, indicating potential implications in disease mechanisms \cite{wu2023reduced}. So a connection between this and our findings give us a new perspective about how having a cytosine after the CpG site might effect more our aging process than having A/T.
% This approach provides profound insights into how biological information impacts aging. 

Currently, most research is fragmented in the sense that different types of clocks, such as, transcriptomic clock, epigentic clock etc. are using different phases of the same process as input. But creating proteins from gene expression regulated by methylation is a connected process. Thus, if all these information about an individual were available, we could utilize them leveraging the power of GraphAge to understand the inner mechanisms of this complex process comprehensively. So, GraphAge can potentially be the first benchmark for a multimodal model which can incorporate (and meaningfully handle) various information about CpG sites, gene expression, DNA sequence mutation etc. as nodes and interaction among these as edges. This could bring us closer to understanding the true nature of aging. 

We also like to remark that our method of CpG site selection was based on AltumAge \cite{lapierre2022pan}. But these sites do not include any sites from X and Y chromosomes, inclusion whereof may unearth further insight, particularly with respect to sex. Also we would point out that our environment was very constraint so the peak of it's potential was not achieved. Thus, if we can lift the restriction in both dimensions- compute power and data- GraphAge promises to unleash its true potential in taking us as close as we can get to the fountain of youth.

%with unrestricted co more resources and build a unified dataset containing all the information (such as gene expression, DNA mutation etc.) and encode their relationships as edges we will be able to harness the true power of this model to build a true representation of biological aging. And by achieving this, I dare to see that, we will be one step closer to achieving the long awaited fountain of youth.

%So the platforms should be standardized so that research can be done in a more systematic way. We would also like to say that our model is open for more inclusion of node attributes and edge attributes which will further improve our model. 

\section{Methods}\label{sec:method}

\subsection{Data collection}\label{dataset}
Datasets from NCBI \cite{ncbi} and EBI \cite{ebi} were used. Datasets were in two different databases - Gene Expression Omnibus (GEO) and Array Express. Because Gene expression patterns are tissue dependent, a tissue type (blood) was selected. They contain both healthy (3707) and unhealthy samples (624). Only healthy data points were used for model training and testing. Among the unhealthy samples, we have postmenopausal women with ovarian cancer (266), schizophrenia (326) and osteoporosis (32).

\subsection{Data preprocessing}\label{data-process}
% Data points with methylation values outside -0.5 to 1.5 were discarded.

Missing values were imputed with K-nearest neighbors (with $K=5$). The dataset were created using two different technologies. Therefore, they were not compatible. To fix this issue, a normalization method called Beta-mixture quantile normalization (BMIQ) was used \cite{teschendorff2013beta}.

%\subsection{Formulation of the Graph}
%In our formulation we needed to convert tabular data into graphs and every graph has nodes and edges. 
%Our nodes are the CpG sites and edges are their relationships. 
%\subsubsection{Selection of CpG Sites and Structural Information}
\subsection{Experimental setup}\label{exp-setup}
 We trained the model on healthy blood tissue datasets of 3707 samples where we first split the data into a test data of 756, i.e., 20\% of the data and the rest where used for 5 fold cross validation. With the final model, we experimented to see the performance of the models in 8 age groups, namely, 0, 0-20, 20-45, 45-55, 55-65, 65-75, 75-80 and 80+ to examine how they performed in various phases of life. Middle ages were more segmented to see how changes occur in those phases of life. %While compiling means of the explanation we used these age groups to understand their interpretation more in depth. %We have used most implementation from pytorch geometric library\cite{pytorch_geometric}.
\subsection{Evaluation Metrics}
As this is a regression task, we used Mean Absolute Error (MAE) and Mean Squared Error (MSE) as our main evaluation metric.
%, and the R\textsuperscript{2} score. 
MSE measures the average of the squares of the errors, giving more weight to larger errors and thus being useful for identifying models that produce significant errors. MAE measures the average magnitude of errors in a set of predictions, without considering their direction, providing a clear indication of the average error and being less sensitive to outliers compared to MSE. Based on prior literature, we also used another metric, age acceleration which is defined as predicted age minus the real age, to understand the direction and value in which the predicted epigenetic age diverts from chronological age.

\subsection{Graph Construction}
%The first question we need answer is which CpG sites should be selected to be turned into nodes. 
In our graph, nodes are the CpG sites and edges are their relationships. Following the structure of AltumAge \cite{lapierre2022pan}, we selected CpG sites that were common across all types of methylation data platforms. The structural information for these CpG sites was obtained from the public website NCBI GEO's supplementary file \cite{ncbi_geo}.
Since the information of distance and position is chromosome specific, CpG site information was prepared based on their chromosomal location, ensuring that their positional and distance information were chromosome-specific.

%\subsubsection{Graph Construction}
%A graph comprises two primary components: \textit{nodes} and \textit{edges}. 
For our model, nodes in the graph included the following attributes:
\begin{itemize}
    \item Methylation beta values for each CpG site.
    \item Boolean value for being inside a CpG island (CpG islands are long sequences of Cytosine and Guanine).
    \item Length of the CpG island.
    \item One-hot encoding of the next base pair (three such base pair information available).
    \item Starting base pair position of the island if it is an island, else 0, normalized.
    \item Ending base pair position of the island if it is an island, else 0, normalized.
    \item Normalized distance from transcription start site; null values are given 1 to indicate they are the farthest.
    \item Map\_Info, which is the position of the CpG site for both island and non-island sites, is also normalized and added to the node attribute.
\end{itemize}

The most crucial part of the formulation is encoding the relationships among the edges. Edge features are:
\begin{itemize}
    \item Co-methylation value among two CpG sites, using methylation values of CpG sites from training data.
    \item Boolean value if two CpG sites are on the same chromosome.
    \item Boolean value if two CpG sites are of the same gene.
\end{itemize}
%\subsubsection{Edge Filtering Based on Co-Methylation Values}
We applied three thresholds for edge filtering:
\begin{itemize}
    \item A universal threshold to filter all edges.
    \item A lower threshold for edges on the same chromosome (Secondary Threshold). We have taken this to be 0.2 lower than the universal threshold.
    \item A distance-based threshold within the same chromosome to allow even lower co-methylation edges to be included if they are closer (Tertiary Threshold). We have taken this to be 0.4 lower than the universal threshold.
\end{itemize}

The rationale behind using three thresholds is based on the influence of closer CpG sites, which tend to have a stronger effect on co-methylation \cite{affinito2020nucleotide}. The concept of distance is applicable only within the same chromosomes. This approach allows for more flexibility in selecting edges that are closer but may have slightly lower co-methylation values. These parameters must be chosen carefully, balancing speed and accuracy requirements.

\subsection{Architecture and Training}
Our model comprises of GNN layers followed by a fully connected layer. So after formulating the graph, we used DataLoaders from PyTorch\cite{pytorch_geometric} to create data loaders. The graph was fed into a Graph Neural Network (GNN) in a stochastic manner. We chose the Principal Neighborhood Aggregation (PNA) convolutional  \cite{corso2020principal} layer for our GNN because of multiple aggregators to capture diverse aspects in graphs. After the GNN layer, the output was passed through a fully connected layer to obtain the final age prediction.

We used 5-fold cross-validation to find the best model and then tested the best model on an test dataset. For other models, we followed exactly the same procedure for a fair comparison.

%\subsection{Compiling Explanation}
\subsection{Interpreting the Network}
We used the GNN explainer \cite{ying2019gnnexplainer} to interpret the model. Two types of explanations were generated:
\begin{itemize}
    \item Importance of each node attribute.
    \item Importance of nodes and edges to identify connected subnetworks that work in conjunction.
\end{itemize}

GNN explainer gives us individual sample explanations which we save. We take these individual sample importance of each site and average them based on the age groups that we used for testing. We then create a graphical representation of the average node and edge importance for each age group. We also color-coded these subnetworks for visualization. Importantly, we generated individual explanations for each sample and averaged the explanations across certain age groups to provide a more comprehensive understanding.

\subsection{Color Code for Graph}\label{Color code}

We employed vizgraph\cite{graphviz} for constructing visualizations:
\begin{itemize}
    \item \textbf{Red node}: hypomethylating CpG sites. With age, the methylation value decreases, and methylation upregulates gene expression.
    \item \textbf{Blue node}: hypermethylating CpG sites. With age, the methylation value increases, and methylation downregulates gene expression.
    \item \textbf{Green circle around node}: the importance of that particular CpG site increases with age.
    \item \textbf{Yellow circle around node}: the importance of that particular CpG site decreases with age.
\end{itemize}

\subsection{Graph Annotation}
\label{graph annotation}
We annotated the graph for better understanding
\subsubsection{Node annotation}
The nodes contain:
\begin{itemize}
    \item CpG\_site name
    \item gene name
    \item distance from Transcription start site
    \item importance of the CpG\_site from GNN explainer
    \item  hyper- or hypomethylation and it's value
    \item Chromosome number
\end{itemize}
\subsubsection{Edge annotation}
\begin{itemize}
    \item comethylation value
    \item importance of the edge i.e. relationship from GNN explainer
\end{itemize}

\subsection{Importance Analysis}

After using the GNN explainer, we analyzed the importance of node attributes by averaging them across all samples. We also plotted the importance of these attributes over different age groups to observe their temporal patterns. For node and edge importance, we divided all samples into various age groups and averaged the importance within each group. We then eliminated nodes with zero importance and edges with low importance. The resulting subnetworks were visualized using Graphviz. We also created individual visualizations for subnetworks that contained a certain number of CpG sites. Additionally, we identified genes from these subnetworks and categorized them based on whether they were hypo- or hypermethylating. We performed Enrichr analysis \cite{chen2013enrichr,kuleshov2016enrichr,xie2021gene} on these two groups of genes separately.

Furthermore, we used linear regression to find the slope of each node's importance with respect to age. We identified the top 10 nodes with the most positive slopes as upward trending and the top 10 nodes with the most negative slopes as downward trending. Nodes with small slope values were considered to have no significant trend.

% \subsection{Graph Traversal}

% We performed graph traversal to find connections of one gene with all its related subnetworks to find pathways related to their behavior of being hypomethylating and hypermethylating. We first filtered edges based on their importance, followed by Depth First Search (DFS) to find all connected subnetworks. Finally, we used node filtering based on the importance of nodes.
\subsection{Code, Environment and Availability}
We used kaggle free tier version using the GPU P100. We have used most implementation from pytorch geometric library \cite{pytorch_geometric}.
%We used Kaggle free tier version as our environment. 
All the datasets and other information are all present in the following notebooks.

Our code for predicting and compiling explanation:
\url{https://www.kaggle.com/code/sakibahmed91/graphage-final-version-for-result-and-explanation}

For analysing the explanation : \url{https://www.kaggle.com/code/salehsakibahmed/graph-age-exaplanation-analysis}

All codes and result compilation are available here: \url{https://github.com/bojack-horseman91/GraphAge}

\backmatter

\bibliography{sn-bibliography}% common bib file

\section*{Supplementary Materials}
\label{Supplementary}

%We can see from Table \ref{table_with_results} that our model's performance is quite comparable with the rest of the models. 

\subsection*{Effect of change of  threshold for filtering edges}
\begin{figure}[htbp!]
    \centering
    \begin{subfigure}[b]{0.4\textwidth}
        \centering
        \includegraphics[width=\textwidth]{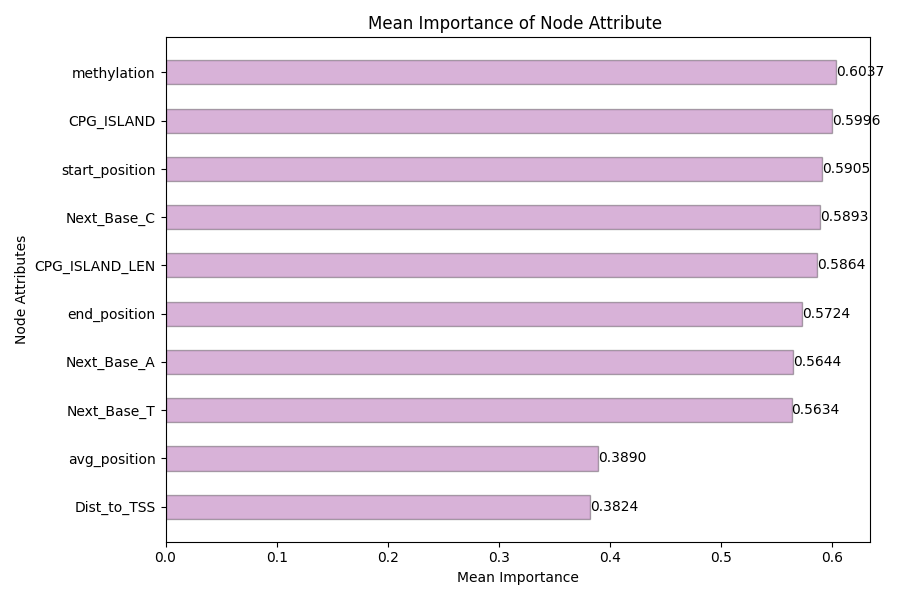}
        \caption{}
        \label{node_mean_importance_0.75}
    \end{subfigure}
    \begin{subfigure}[b]{0.4\textwidth}
        \centering
        \includegraphics[width=\textwidth]{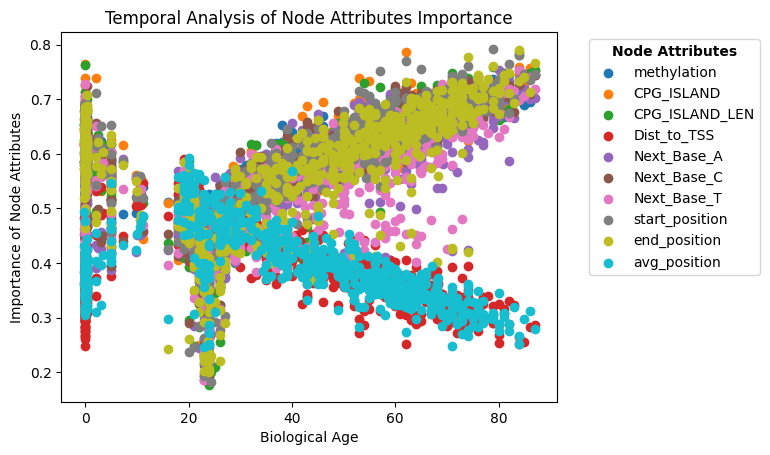}
        \caption{}
        \label{temporal_analysis_of_node_importance_0.75}
    \end{subfigure}
    \caption{\textbf{a.} For threshold 0.75, secondary\_threshold 0.73 and tertiary\_threshold of 0.71 we trained GNN explainer for 150 epoch. And by using the GNN explainer, we determined the importance of each node attribute for every individual sample. We saved these results in a dataset and analyzed them in a new notebook. The importances were then averaged to provide a broader understanding of which attributes are most influential. We observe that the methylation beta value is the most important factor in predicting age. After that we see some interesting things. First is information about CpG islands and next base pairs are very important in predicting age. We can see that the start position of the island and whether the node (i.e., CpG site) is a CpG Island is quite significant. This is followed by whether the next base pair is C or not. Additionally, the length of the island is also prominent, indicating that being a CpG Island has a significant effect on the aging process. We can also see the effect of other next base pair in the CpG site. Conversely, the distance to the transcription start site (Dist\_TSS) and the average position of the site are less influential. Therefore, the location of a CpG Island and its length, along with the starting position of the island and the next base pair, play a crucial role in the aging process.\textbf{b.} We also performed a temporal analysis by plotting the importance with respect to chronological age. We observe that the distance to the transcription start site (Dist\_TSS) and Map\_Info (the exact position of the site) see a slight increase up to the 20s, whereas the rest decrease up to the 20s. Afterwards, they do the opposite: Dist\_TSS and Map\_Info (exact position) decrease in importance, while the rest have their importance increased.}
    \label{node_analysis 0.75}
\end{figure}

\begin{figure}[htbp!]
    \centering
    \includegraphics[width=0.5\linewidth]{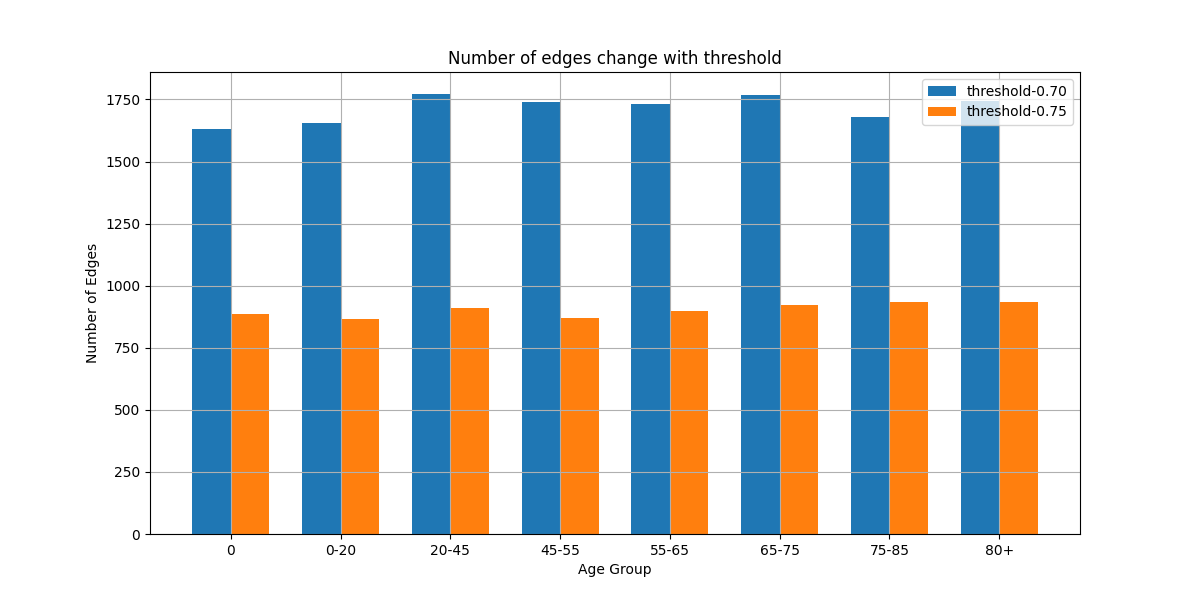}
    \caption{By decreasing the threshold from 0.75 to 0.7 we see an increase of on average increase of 812.75 (90.162\%) in the number of edges after filtering in the Methylation Regulated Network i.e. graph after filtering less important (edge importance less than 0.1) edges and zero important nodes i.e. CpG sites. So by decreasing the threshold we can get more relationships for better understanding the complexities of aging.}
    \label{fig:Change of edges with threshold}
\end{figure}

Our paper includes all results from using threshold 0.7 but in Fig. \ref{node_analysis 0.75} we show the results for threshold 0.75, secondary\_threshold 0.73 and tertiary\_threshold of 0.71. One other difference is that the number of epoch to train GNN explainer in this case was 150 while in the explanation of threshold 0.7 epoch was 100 due to longer training time. %But from Fig. \ref{node analysis} and Fig. \ref{node_analysis 0.75} it becomes evident that the importance of being a CpG island is very significant. Also among the next base pair , Base pair C is most influential. Using this observation we gain a lot of key insights.
As can be observed from the figure, particularly in contrast with Fig. \ref{temporal analysis of node importance}, although node attribute importance remains quite similar the edge importance scores change. We can see from Fig. \ref{fig:Change of edges with threshold} that by decreasing the threshold from 0.75 to 0.7, on an average for each age group, we see an increase of 812.75 (90.162\%) in the number of edges. 
%after filtering in the Methylation Regulated Network i.e. graph after filtering less important (edge importance less than 0.1) edges and zero important nodes i.e. CpG sites. 
So by decreasing the threshold we are encoding more interactions/relationships into the graph structure thereby increasing the potential for a more comprehensive understanding of the complexities of aging while decoding (i.e., interpreting) it. Table \ref{table_with_results} provides a detailed comparison in terms of MAE among GraphAge, AltumAge, and Horvath’s model including results for multiple threshold values of GraphAge. Also, Table \ref{age_group_comp_table} presents a comparison between GraphAge and AltumAge in different age groups.

\subsection*{More on the identified genes}
\label{sec:age_related_further_analysis}
% \subsubsection*{Connection with mice} 
There are multiple studies in the literature on mice that linked the identified important genes with aging. 
%In our studies of aging-related pathways, we identified genes whose CpG sites had the highest importance scores and found various studies connecting them with aging. We now also present some research conducted on mice that connects two of the most frequently mentioned genes, TSC1 and PDPK1, in the mTOR and LEP in AMPK pathways respectively, with aging.
TSC1, a tumor suppressor gene \cite{ncbi_TSC1}, is shown to be associated with accelerated retinal aging in knockout mice \cite{rao2021mtorc1}. Moderate lifelong overexpression of TSC1 has been demonstrated to improve health and survival in mice \cite{zhang2017moderate}. PDPK1 signaling has great influence on determining the reproductive aging and the length of reproductive life in females \cite{reddy2009pdk1}. On the other hand, aging affects leptin (LEP) actions differently. There is a decline in hypermetabolic responses but increased sensitivity in lean rats. Older rats are more prone to obesity-induced leptin resistance, while calorie restriction enhances leptin responsiveness, particularly in older rats \cite{balasko2014leptin}.

% Research indicates that TSC1, a tumor suppressor gene \cite{ncbi_TSC1}, is associated with accelerated retinal aging in knockout mice \cite{rao2021mtorc1}. Additionally, moderate lifelong overexpression of TSC1 has been shown to improve health and survival in mice \cite{zhang2017moderate}. Studies also reveal that Aging affects leptin (LEP) actions differently, with a decline in hypermetabolic responses but increased sensitivity in lean rats. Old rats are more prone to obesity-induced leptin resistance, while calorie restriction enhances leptin responsiveness, particularly in older rats \cite{balasko2014leptin}.

\subsection*{More on the Dataset}
These are all the dataset that we used in our entire process. For clarity of user we have added a small description for each of the dataset. 
\begin{itemize}[label={}]
    \item \textbf{E-GEOD-51388:} Blood samples taken longitudinally from monozygotic twins. The platform used was Illumina’s Infinium 450k Human Methylation Beadchip.
    \item \textbf{E-GEOD-52588:} Blood samples from subjects with or without Down syndrome. The platform used was Illumina’s Infinium 450k Human Methylation Beadchip.
    \item \textbf{E-GEOD-53128:} Whole blood samples. The platform used was Illumina’s Infinium 450k Human Methylation Beadchip.
    \item \textbf{E-GEOD-53740:} Blood samples from progressive supranuclear palsy patients, frontotemporal dementia patients, and healthy controls. The platform used was Illumina’s Infinium 450k Human Methylation Beadchip.
    \item \textbf{E-GEOD-54399:} Cord blood and placenta. All ages were encoded as zero age. The platform used was Illumina’s Infinium 450k Human Methylation Beadchip.
    \item \textbf{E-GEOD-54690:} Blood samples from subjects with or without dietary flavanol supplementation. The platform used was Illumina’s Infinium 450k Human Methylation Beadchip.
    \item \textbf{E-GEOD-56553:} Peripheral mononuclear blood cells of asthmatic patients. The platform used was Illumina’s Infinium 450k Human Methylation Beadchip.
    \item \textbf{E-GEOD-57484:} Whole blood samples from normal and obese children. The platform used was Illumina’s Infinium 27k Human Methylation Beadchip.
    \item \textbf{E-GEOD-58045:} Whole blood samples. The platform used was Illumina’s Infinium 27k Human Methylation Beadchip.
    \item \textbf{E-GEOD-59509:} Whole blood, saliva, menstrual blood, vaginal swab, and semen samples. The platform used was Illumina’s Infinium 450k Human Methylation Beadchip.
    \item \textbf{E-GEOD-59592:} Blood from infants exposed to varying degrees of aflatoxin B1. The platform used was Illumina’s Infinium 450k Human Methylation Beadchip.
    \item \textbf{E-GEOD-62219:} Longitudinal peripheral blood leukocyte samples from infants. The platform used was Illumina’s Infinium 450k Human Methylation Beadchip.
    \item \textbf{E-GEOD-64495:} Whole blood from subjects with or without developmental disorder syndrome X. Non-healthy patient samples were separated for further analysis and not included in training, validation, or testing. The platform used was Illumina’s Infinium 450k Human Methylation Beadchip.
    \item \textbf{E-GEOD-64940:} Cord blood samples from newborns. All samples were encoded as zero age. The platform used was Illumina’s Infinium 27k Human Methylation Beadchip.
    \item \textbf{E-GEOD-65638:} Whole blood samples from twins. The platform used was Illumina’s Infinium 450k Human Methylation Beadchip.
    \item \textbf{E-GEOD-67444:} Neonatal blood samples. All ages were encoded as zero age. The platform used was Illumina’s Infinium 450k Human Methylation Beadchip.
    \item \textbf{E-GEOD-67705:} Blood samples from HIV+ and HIV- subjects. The platform used was Illumina’s Infinium 450k Human Methylation Beadchip.
    \item \textbf{E-GEOD-71245:} Blood samples from different types of blood cells. The platform used was Illumina’s Infinium 450k Human Methylation Beadchip.
    \item \textbf{E-GEOD-71955:} CD4+ and CD8+ T-cell samples from subjects with Graves’ disease or healthy controls. The platform used was Illumina’s Infinium 450k Human Methylation Beadchip.
    \item \textbf{E-GEOD-72338:} Neutrophils and monocytes from patients with tuberculosis and household controls. The platform used was Illumina’s Infinium 450k Human Methylation Beadchip.
    \item \textbf{E-GEOD-77445:} Whole blood samples from subjects with different stress levels. The platform used was Illumina’s Infinium 450k Human Methylation Beadchip.
    \item \textbf{E-GEOD-79056:} Cord blood samples. The platform used was Illumina’s Infinium 450k Human Methylation Beadchip.
    \item \textbf{E-GEOD-83334:} Whole blood and cord blood from newborns and infants measured longitudinally. The platform used was Illumina’s Infinium 450k Human Methylation Beadchip.
    \item \textbf{E-MTAB-2344:} White blood cell samples in patients with stroke and/or obesity and healthy controls. The platform used was Illumina’s Infinium 27k Human Methylation Beadchip.
    \item \textbf{E-MTAB-2372:} White blood cell samples from obese patients subject to two different diets. The platform used was Illumina’s Infinium 450k Human Methylation Beadchip.
    \item \textbf{GSE19711:} Whole blood samples from patients with or without ovarian cancer. The platform used was Illumina’s Infinium 27k Human Methylation Beadchip.
    \item \textbf{GSE20236:} Whole blood samples. The platform used was Illumina’s Infinium 27k Human Methylation Beadchip.
    \item \textbf{GSE20242:} CD4+ T-cells and CD14+ monocytes samples. The platform used was Illumina’s Infinium 27k Human Methylation Beadchip.
    \item \textbf{GSE34257:} Blood cord samples and whole blood samples from 9-month-old infants. The platform used was Illumina’s Infinium 27k Human Methylation Beadchip.
    \item \textbf{GSE36642:} Blood mononuclear cell, human umbilical vascular endothelial cell, and placenta samples from monozygotic and dizygotic twins. The platform used was Illumina’s Infinium 27k Human Methylation Beadchip.
    \item \textbf{GSE37008:} Peripheral blood mononuclear cell samples. The platform used was Illumina’s Infinium 27k Human Methylation Beadchip.
    \item \textbf{GSE41037:} Whole blood samples in schizophrenia patients and healthy subjects. The platform used was Illumina’s Infinium 27k Human Methylation Beadchip.
    \item \textbf{GSE49904:} Blood buffy coat samples. The platform used was Illumina’s Infinium 27k Human Methylation Beadchip.
    \item \textbf{GSE56606:} CD14+ monocyte samples from monozygotic twins discordant for type 1 diabetes. The platform used was Illumina’s Infinium 27k Human Methylation Beadchip.
    \item \textbf{GSE57285:} Whole blood samples from women with BRCA1 wild-type or mutants, and with or without breast cancer. The platform used was Illumina’s Infinium 27k Human Methylation Beadchip.
    \item \textbf{GSE69176:} Umbilical cord blood samples from newborns. The platform used was Illumina’s Infinium 450k Human Methylation Beadchip.
    \item \textbf{GSE99624:} Whole blood samples from osteoporotic and healthy control patients. The platform used was Illumina’s Infinium 450k Human Methylation Beadchip.
\end{itemize}
\FloatBarrier
\begin{table}[htbp!]
\begin{center}
\resizebox{\textwidth}{!}{
\begin{tabular}{ |c|c|c|c|c|c|c|c|c|} 
\hline
%trained on our dataset
%off the shelf model used without raining
\shortstack{\textbf{Tissue} \\ \textbf{name}} & \shortstack{\textbf{AltumAge} \\ \textbf{(trained} \\ \textbf{on} \\ \textbf{our dataset)}} & \shortstack{\textbf{Horvath's} \\ \textbf{clock} \\ \textbf{(trained} \\ \textbf{on} \\ \textbf{our dataset)}} & \shortstack{\textbf{Horvath} \\ \textbf{(off the} \\ \textbf{shelf model)}\\-used without training} & \shortstack{\textbf{AltumAge} \\ \textbf{(off the} \\ \textbf{shelf model)}\\-used without training} & \shortstack{\textbf{GraphAge} \\ \textbf{(Threshold:} \\ \textbf{0.80,} \\ \textbf{Secondary:} \\ \textbf{0.79,} \\ \textbf{Tertiary:} \\ \textbf{0.76)}} & \shortstack{\textbf{GraphAge} \\ \textbf{(Threshold:} \\ \textbf{0.75,} \\ \textbf{Secondary:} \\ \textbf{0.74,} \\ \textbf{Tertiary:} \\ \textbf{0.71)}} & \shortstack{\textbf{GraphAge} \\ \textbf{(Threshold:} \\ \textbf{0.7,} \\ \textbf{Secondary:} \\ \textbf{0.68,} \\ \textbf{Tertiary:} \\ \textbf{0.64)}}\\
\hline
\shortstack{Blood \\ \shortstack{(Sample count: \\Train: \\ 2361, \\ Validation: \\ 590, \\ Test: \\ 756)}} & 3.29 & 4.075 & 5.317 & 3.71 & 3.5 & 3.35 & 3.20 \\
\hline
\end{tabular}
}
\end{center}
\caption{The Table provides a detailed comparison of the Mean Absolute Error (MAE) values for GraphAge, AltumAge, and Horvath's model including results for multiple threshold values for GraphAge. The results presented here are based on the test set. Here, ``off the self model" means that we did not train them, rather took the model weights from the literature and used those directly on our test set.}
\label{table_with_results}
\end{table}

% Performance table

\begin{table}[h!]
\centering
\begin{tabular}{|c | c | c | c | c | c | c | c | c | c |}
\hline
\shortstack{Age \\ Group} & \shortstack{sex} & \shortstack{GraphAge \\ MAE} & \shortstack{GraphAge \\ MSE} & \shortstack{number of\\ samples} & \shortstack{AltumAge \\ MAE} & \shortstack{AltumAge \\ MSE} \\
\hline
0 & M & 0.20 & 1.53  & 83 & 0.31 & 4.18 \\
\hline
0 & F & 0.19 & 0.54  & 68 & 0.25 & 0.72 \\
\hline
0-20 & M & 1.79 & 10.24  & 34 & 1.76 & 6.28  \\
\hline
0-20 & F & 2.60 & 15.64  & 38 & 2.56 & 14.00  \\
\hline
20-45 & M & 3.20 & 21.95  & 101 & 3.30 & 26.15  \\
\hline
20-45 & F & 3.36 & 21.02  & 109 & 4.06 & 29.30  \\
\hline
45-55 & M & 3.32 & 17.04  & 24 & 2.97 & 15.58  \\
\hline
45-55 & F & 3.89 & 24.94  & 63 & 4.04 & 31.63 \\
\hline
55-65 & M & 4.16 & 28.12  & 19 & 5.21 & 40.36 \\
\hline
55-65 & F & 4.44 & 43.95  & 101 & 4.54 & 49.29 \\
\hline
65-75 & M & 6.76 & 68.39  & 16 & 6.16 & 48.58 \\
\hline
65-75 & F & 5.49 & 53.09  & 70 & 4.42 & 49.49 \\
\hline
75-80 & M & 5.74 & 38.40  & 2 & 7.78 & 65.97 \\
\hline
75-80 & F & 6.34 & 50.69  & 18 & 6.67 & 55.30\\
\hline
80+ & M & 5.00 & 30.05  & 4 & 6.89 & 53.92 \\
\hline
80+ & F & 9.48 & 103.70  & 6 & 10.26 & 121.50 \\
\hline
\end{tabular}
\caption{Age wise comparison of performance between GraphAge and AltumAge on the blood dataset.}
\label{age_group_comp_table}
\end{table}

% \begin{figure}
%     \centering
%     \includegraphics[width=0.5\linewidth]{images/flank.png}
%     \caption{This is a borrowed image from \cite{gao2020comprehensive}. The Weblogos representing the 50–200 most preferred NNNCGNNN methylation sites by mDNMT3A and mDNMT3B reveal distinct nucleotide preferences at the position immediately following the CpG site. For mDNMT3A, the most preferred base pair at this position is Cytosine, while for mDNMT3B, Guanine is favored. This indicates a specific sequence preference by these methyltransferases in their methylation patterns.}
%     \label{fig:borrowed flank image}
% \end{figure}
\begin{table}[h!]
    \centering
    \begin{tabular}{|l|p{10cm}|}
        \hline
        \textbf{Variable} & \textbf{Short Description} \\
        \hline
        CPG\_ISLAND & Identifies whether a CpG site is located in a CpG Island. \\
        \hline
        CPG\_ISLAND\_LEN & Indicates the length of the island. \\
        \hline
        start\_pos\_of\_ISLAND & Represents the start position of the CpG island. \\
        \hline
        end\_pos\_of\_ISLAND & Represents the end position of the CpG island. \\
        \hline
        Dist\_TSS & Information of Transcription Start Site (TSS) distance of all sites including Islands. \\
        \hline
        Map\_Info & The exact position of all sites. \\
        \hline
        Next\_Base\_A & Next i.e. adjacent base pair after CpG site is A. \\
        \hline
        Next\_Base\_T & Next i.e. adjacent base pair after CpG site is T. \\
        \hline
        Next\_Base\_C & Next i.e. adjacent base pair after CpG site is C. \\
        \hline
    \end{tabular}
    \caption{Description of Node attributes variables}
    \label{Table: Node attribute variable des}
\end{table}

\end{document}